\documentclass[11pt]{article}

\usepackage{sectsty}
\usepackage{graphicx}
\usepackage{url}
\usepackage{hyperref}
\usepackage{authblk}
\usepackage[table]{xcolor}
\usepackage{multirow}
\usepackage{hhline}
\usepackage{siunitx}
\usepackage{booktabs}
\usepackage{longtable}
\usepackage{tablefootnote}
\usepackage{tabularray}\UseTblrLibrary{booktabs}

\usepackage{pdflscape}

\usepackage{lineno}

\newcommand{\bb}[1]{\textbf{#1}}
\renewcommand{\it}[1]{\textit{#1}}
\newcommand{\bi}[1]{\bb{\it{#1}}}
\newcommand{\txt}[1]{\texttt{#1}}

\newcommand{\Parti}[1]{Participant \texttt{#1}}
\newcommand{\etal}{et al.}
\providecommand{\keywords}[1]{\it{\bb{Keywords:} #1}}

\graphicspath{{./figures}}

\title{How do practitioners gain confidence in assurance cases?}

\author[1]{Simon Diemert}
\author[1]{Caleb Shortt}
\author[1]{Jens H. Weber}
\affil[1]{Department of Computer Science, University of Victoria, Canada}
\affil[ ]{\{\tt{sdiemert,cshortt,jens}\}@uvic.ca}

\date{}

\begin{document}
\maketitle

\begin{abstract}
\noindent \bb{CONTEXT:} Assurance Cases (ACs) are prepared to argue that the system's desired quality attributes (e.g., safety or security) are satisfied. While there is strong adoption of ACs, practitioners are often left asking an important question: \it{are we confident that the claims made by the case are true?} While many confidence assessment methods (CAMs) exist, little is known about the use of these methods in practice.

~\\
\noindent\bb{OBJECTIVE:} Develop an understanding of the current state of practice for AC confidence assessment: what methods are used in practice and what barriers exist for their use?

~\\
\noindent\bb{METHOD:} Structured interviews were performed with practitioners with experience contributing to real-world ACs. Open-coding was performed on transcripts. A description of the current state of AC practice and future considerations for researchers was synthesized from the results.

~\\
\noindent\bb{RESULTS:} A total of $n=19$ practitioners were interviewed. The most common CAMs were (peer-)review of ACs, dialectic reasoning (``defeaters''), and comparing against checklists. Participants preferred qualitative methods and expressed concerns about quantitative CAMs. Barriers to using CAMs included additional work, inadequate guidance, subjectivity and interpretation of results, and trustworthiness of methods.

~\\
\noindent\bb{CONCLUSION:} While many CAMs are described in the literature there is a gap between the proposed methods and needs of practitioners. Researchers working in this area should consider the need to: \it{connect} CAMs to established practices, use CAMs to \it{communicate} with interest holders, \it{crystallize} the details of CAM application, \it{curate} accessible guidance, and \it{confirm} that methods are trustworthy.

~\\
\noindent\keywords{Assurance Cases, Safety Cases, Confidence Assessment, Practitioner Interviews}
\end{abstract}

\tableofcontents

\section{Introduction}\label{sec:introduction}

Critical systems are those where failure can lead to significant losses in terms of economic/financial harm, environmental damage, injury to humans, or loss of life, and are referred to as business-, mission-, security-, or safety-critical systems. The challenge of assuring these systems has increased as novel technologies and capabilities (e.g., artificial intelligence, autonomy, self-adaptation) are relied upon to realize critical functions. For instance, autonomous vehicles depend on artificial intelligence (mainly machine learning) methods to perceive their environment and detect nearby agents, and perception system failures have lead to severe consequences \cite{kohli2020,koopman2024}. Significant effort is expended to develop these systems and avoid such loss events.

Assurance Cases (ACs) contain arguments, supported by evidence such as analyses or verification results, that a system will satisfy essential quality attributes within its defined operating environment \cite{gsn}. ACs are sometimes also referred to as safety cases or security cases to illustrate the primary quality attribute they are assuring. ACs are required for compliance with technical standards and regulations in a range of industries, including: automotive \cite{iso26262,ul4600,iso21434}, rail signalling and control \cite{en50126}, nuclear power systems \cite{onr2019}, defence \cite{defstan56}, and oil \& gas \cite{offshore2005}. There is also interest in preparing ACs for medical devices \cite{aami2019} and commercial aviation systems \cite{holloway2019,wasson2022}. ACs are especially useful for assuring novel systems, where established approaches to systems assurance are less effective or difficult to apply. For instance, a group of leading experts recently (re-)affirmed the role of ACs in assuring autonomous vehicles \cite{dagstuhl2024}. ACs are also increasingly recognized as ``live'' models of assurance that evolve alongside their system, particularly in the fields of autonomous and (self-)adaptive systems; such ACs are referred to as \it{dynamic assurance cases} \cite{denney2015,calinescu2018,asaadi2020}.

While there is strong adoption of ACs as a means of capturing arguments and supporting evidence, not all arguments or evidence are equal. For instance, arguments that are convincing in one context might not hold for another, or evidence might be produced through untrustworthy means. Moreover, for dynamic ACs, the evidence might change over time, potentially invalidating claims that were previously thought to be true. So, when preparing or managing an AC, an important question arises: \it{are we confident that the claims made in an AC are true?} 

Confidence in ACs has been the subject of significant research for nearly two decades. Many authors have developed methods for assessing confidence in ACs \cite{hawkins2011,hobbs2012,goodenough2015,holloway2021,idmessaoud2024}. Throughout the rest of this paper, we collectively refer to these (and other procedures, techniques, approaches, and activities intended to assess confidence in an AC) as \it{confidence assessment methods} (CAMs). While the development of each CAM was likely informed by its creator's experience interacting with real-world systems and problems, there is a deficit of empirical research investigating if practitioners use these methods to prepare ACs for real-world systems and whether they find them useful. A consequence of this is that CAMs are developed or extended without a complete understanding of the needs of their end-users: the practitioners creating ACs for real-world critical systems.

\subsection{Summary of Contribution}

This paper reports on a series of structured interviews with AC practitioners. The overall objective of this study was to develop an understanding of practitioners' use of CAMs while preparing or managing ACs for real-world systems, including the specific CAMs used and barriers to their use. In other words, our aim is to understand how practitioners gain confidence in the claim(s) made in ACs. This study adopted a grounded-theory perspective and used open coding to analyze the interview transcripts and extract themes and concepts. The result is a description of the current state of practice for the study's sample of practitioners for ACs and CAMs. Additionally, the findings are synthesized into five considerations for developing (or extending) CAMs. While other authors have published empirical studies (interviews, questionnaires) in the area of ACs \cite{torner2008,nair2015,doss2016,cheng2018,almendra2022}, this is the first empirical work investigating the real-world use of CAMs by practitioners.

\subsection{Paper Structure}

The remainder of this paper is structured as follows. To begin with, Section \ref{sec:background} surveys existing CAMs from the literature. Next, the study's method is described in Section \ref{sec:method} and the results follow in Section \ref{sec:results}. Then, we review similar empirical studies on ACs and compare our results in Section \ref{sec:related}. Finally, we present a summary of our study's findings and considerations for researchers developing CAMs in Section \ref{sec:discussion} and Section \ref{sec:conclusion} closes with concluding remarks.

\section{Survey of Confidence Assessment Methods}\label{sec:background}

This section surveys CAMs that are described in the literature. Broadly speaking, confidence assessment methods can be categorized as either qualitative, quantitative, or mixed (both qualitative and quantitative). It is assumed that readers are familiar with ACs in general, including their role in a safety engineering lifecycle and methods/notations for expressing them (e.g., the Goal Structuring Notation (GSN) \cite{gsn,kelly}).

It is worth noting that inclusion of a method in this section does not mean that practitioners in the study reported its use. Several methods are included in this section for completeness but were not mentioned by practitioners.

\subsection{Qualitative Methods}

Qualitative CAMs focus on assessing or increasing confidence based on qualities of the AC (or lack thereof) and do not assign numerical valuations to describe one's level of confidence.

\paragraph{Assurance Claim Points.} Hawkins \etal{} separate an AC's argument into two independent, but connected structures \cite{hawkins2011}. The main argument captures the fundamental rationale for why a system is assured. The secondary confidence argument provides an additional layer of argumentation to describe why authors believe the claims in the main argument. The two arguments are connected through Assurance Claim Points (ACPs). Hawkins \etal{} use GSN to describe both arguments and denote the claim points where the arguments intersect as shaded rectangles attached to the links (edges) in the GSN structure. Their motivation for proposing the ACP method is to simplify the main assurance argument such that it can be easily understood by readers, and detailed confidence arguments are still available for those that wish to examine them. Hawkins \etal{} recently described a variant of ACPs, which they call \it{operational claim points} (OCPs), aimed at capturing arguments for why an AC remains valid while the system is in operation \cite{fenn2024}.

\paragraph{Dialectic Argumentation.} ACs are vulnerable to confirmation bias where authors favour arguments and evidence they already believe to be true without considering alternative perspectives which has contributed to severe accidents \cite{leveson2011,nimrod}. Dialectic argumentation is a method where ``challenges'', ``defeaters'', ``counter-claims'', or ``weakeners'' representing doubt or uncertainty in either the argument or supporting evidence are raised against an argument's positive claims \cite{shahandashti2024}. Dialectic argumentation has been incorporated into several structured AC notations, including Toulmin's notation \cite{toulmin}, GSN \cite{gsn,acwg2021}, Eliminative Argumentation (EA) \cite{goodenough2015}, Friendly Argument Notation \cite{fan}, Assurance 2.0 \cite{bloomfield2023}, and the structured assurance case meta-model (SACM) \cite{sacm}. Throughout the remainder of this paper we use the term ``defeater'' to refer to dialectic argumentation. There are several reports in the literature of defeaters being used by practitioners to address confirmation bias for real-world systems\cite{hobbs2019,diemert2020,millet2023,ryan2024}. Once expressed in an argument, a defeater can be ``resolved'' through further argumentation or by providing additional evidence. However, there are differing perspectives on how to interpret resolved defeaters and the implications for confidence in the AC \cite{diemert2024}. The ``deductive perspective'' suggests that resolving a defeater should strictly return the argument to a neutral state, as if the defeater had never existed. Alternatively, the ``defeasible reasoning'' perspective advocates that some additional credit should be earned for the extra work required to both identify and resolve the defeater, and thus have a net positive effect on confidence \cite{goodenough2013}.

\paragraph{iTest.} Holloway and Wasson propose the iTest method for qualitatively evaluating arguments \cite{holloway2021}. While they have described their method in terms of the Friendly Argument Notation (FAN) \cite{fan}, the underlying concepts are broadly applicable to other notations and approaches for expressing arguments. The iTest method is performed iteratively, where compound arguments have their atomic sub-arguments isolated and interrogated. Each atomic argument is assessed using the SPRY mnemonic as a guide: ``is the \bb{S}yntax proper?'', ``are the \bb{P}remises acceptable?'', ``is the \bb{R}easoning acceptable?'', and ``is saying \bb{Y}es to the conclusion justified?'' \cite{holloway2021}. iTest has seen early use in the context of evaluating whether aerospace systems possess desirable ``over-arching properties'' \cite{wasson2022}.

\subsection{Quantitative Methods}

\newcommand{\josang}{J{\o}sang}

Quantitative CAMs assess confidence by assigning numerical valuations of confidence to aspects of the argument and then use an algorithm to propagate the valuations through the argument's structure. The result is one or more numbers describing confidence in the top-level claim of the argument. The following decomposes quantitative CAMs into ``families'' of methods according to the theoretical basis of the method.

\paragraph{Bayesian Networks.} Several proposals to use Bayesian Networks (BNs) to assess confidence exist \cite{denney2011,hobbs2012,zhao2012,oh2022}. This family of methods typically models an argument structure as a Bayesian Network (BN) and then assigns a probabilistic valuation to the leaves of the argument structure which is propagated through the BN using the laws of probability. Sometimes the user of the method must select additional parameters (e.g., ``weights'') to adjust the propagation through the argument. For instance, Hobbs and Lloyd's method requires that analysts provide a ``belief'' valuation (a subjective probability in $[0,1]$) for each leaf node of the argument, an additional link parameter for each parent-child relationship, a combinator parameter (i.e., ``AND'' or ``OR''), and a leakage parameter for each parent node in the argument. Once all inputs and parameters are given the method uses a BN to compute the degree of belief (as a probability) in the argument's top-level claim \cite{hobbs2012,diemert2024}.

\paragraph{Dempster-Shafer Theory.} Another family of quantitative CAMs relies on a mathematical framework called Dempster-Shafer Theory (DST) (also called ``evidence theory'' or ``evidential reasoning'') \cite{shafer1976}. These CAMs typically consider confidence as being composed of three concepts: belief, disbelief, and uncertainty that are linked according to a constraint known as \josang{}'s triangle \cite{cyra2011,ayoub2013,wang2019,idmessaoud2024,josang2016}. The most recent method proposal in the DST family is from Idmessaoud \etal{} \cite{idmessaoud2024}. In their approach an analyst provides two valuations to each leaf of the argument: a ``decision'' and the ``confidence'' (uncertainty) in the decision. Then for each parent-child relationship the analyst also provides forward and backward relation parameters describing the relationship between the parent and child. Further, for each parent two more parameters are required to describe the support for the parent in terms of all of its children. Once all inputs and parameters are given, the method uses DST to compute a decision and confidence (uncertainty) for the argument's top-level claim.

\paragraph{Possibilistic Logic.} Possibilistic Logic (PL) is a formalism for reasoning based on ``possible'' and ``necessary'' outcomes using fuzzy set theory \cite{dubois1988,zadeh1965}. Idmessaoud \etal{} applied PL to address several limitations they observed with their DST-based approach, such as confidence amplification and attenuation effects caused by fractions being multiplied together during confidence propagation \cite{idmessaoud2024}. Given that Idmessaoud \etal{}'s PL-based approach is quite new, there are no reports of it being used for real-world systems yet.

\paragraph{Subjective Logic.} Subjective Logic (SL) is another formalism for reasoning under uncertainty developed by \josang{} \cite{josang2016}. Like DST, SL encodes confidence in terms of three quantities: belief, disbelief, and uncertainty, which are called ``opinions''. However, unlike DST, it also provides a mathematical ``bridge'' to express confidence in terms of probability distributions, namely the beta distribution, where the mean value of the distribution encodes belief and the variance captures uncertainty. Since SL is a logic, it provides several operators that can be used to propagate confidence. Several authors have proposed CAMs based on SL \cite{duan2015,yuan2017,herd2024}. The most recent is by Herd \etal{}, which applied SL to an argument fragment focused on autonomous vehicle perception \cite{herd2024}.

\paragraph{Validity of Quantitative Methods.} A survey of quantitative CAMs would be incomplete without mentioning Graydon's and Holloway's work on the validity of these methods \cite{graydon2017}. In summary, Graydon and Holloway selected several qualitative CAMs and attempted to independently reproduce the worked examples in the related publications with minimal success or found that the results the CAMs generate are ``implausible''. This suggests that, even though these methods are built on rigorous mathematical frameworks, additional work is required to validate them for practical use. Graydon and Holloway concluded that quantitative CAMs ``require further validation before they should be recommended as part of the basis for deciding whether an assurance argument justifies fielding a critical system'' \cite{graydon2017}.

\subsection{Mixed Methods}

There are some CAMs that mix both qualitative and quantitative aspects together and so do not fit in either category above.

\paragraph{Baconian Probabilities.} If an argument is prepared using defeaters, then Goodenough \etal{} propose that confidence can be measured using a Baconian Probability (BP) \cite{goodenough2013}. BP's are expressed as $x|y$ where $y$ is the total number of defeaters in the argument and $x$ is the number of resolved defeaters. Note that BPs should not be interpreted as a conventional probability in $[0,1]$, rather they should be read as ``$x$ out of $y$''). BPs are based on the notion of defeasible reasoning, where one's confidence increases as reasons to doubt a claim's truth are enumerated and resolved. However, the magnitude of the increase is not defined. Goodenough \etal{} originally proposed using BPs to evaluate ``confidence maps'', which were the precursor to the notation now referred to as Eliminative Argumentation \cite{goodenough2015}. Some authors have described limitations about the use of BPs as means of quantifying confidence \cite{diemert2020,graydon2016}.

\paragraph{Assurance 2.0.} Bloomfield and Rushby proposed the ``Assurance 2.0'' framework, which considers three complementary perspectives: positive, negative, and residual \cite{bloomfield2023}. Confidence in Assurance 2.0 first comes from building sound arguments in the positive perspective, then ``confirmation measures'' using subjective probability valuations can be used to quantify confidence in the positive perspective. In the negative perspective, the analyst identifies (and aims to resolve) defeaters to the positive argument. The argument is said to be valid if all defeaters are resolved. The residual perspective considers unresolved defeaters, which must be accepted to complete the AC.

\section{Method}\label{sec:method}

This study used a grounded theory methodology with the aim of understanding the current industrial practice for AC confidence assessment where, themes and concepts emerge from analysis of data \cite{glaser1999,easterbrook2008}. Data was collected using structured interviews with AC practitioners and an open-coding analysis was performed on the resulting interview transcripts to identify concepts and themes. The study used a cross-sectional  approach (i.e., a ``snapshot'' in time), and was designed based on the guidance in \cite{seaman1999,kasunic2005,linaker2015}. The remainder of this section describes the methods used for recruitment, interviewing, and data analysis. Since this study involved interviews with human subjects, the methods were reviewed and approved by our institution's research ethics board.

\subsection{Recruitment}

Recruitment used a ``convenience'' sample from the investigators' professional networks. Invitations were sent via email to persons with known industrial or practical experiences building ACs (i.e., practitioners) rather than those with experience conducting research with ACs. Recruitment was targeted to capture a wide range of experiences from different industries, rather than focusing on a single industry. To be included in the study, a participant must have:

\begin{enumerate}
    \item had experience contributing to (authoring, reviewing, assessing, etc.) one or more ACs for real-world industrial system(s);
    \item been a working professional or recently retired from such a role;
    \item been at least 18 years of age (intended to exclude minors); and
    \item been professionally fluent in English to participate in an English language interview.
\end{enumerate}

Invited participants self-applied the inclusion criteria as part of the initial invitation and, if they proceeded to an interview, were screened again by the interviewer with survey questions (see Questions \#1 and \#3 in Appendix \ref{appendix:questions}). Participants that were interviewed, but who did not satisfy the inclusion criteria as determined by screening questions, completed the interview but did not have their interview transcript/data included in the analysis.

\subsection{Interviews}

Structured interviews with synchronous discussion, as opposed to an asynchronous discussion or questionnaire, were chosen as the data gathering method because they allowed in-depth exploration of specific topics to produce richer data for analysis. Interviews were conducted in a structured manner where one interviewer worked through a pre-determined set of questions with the participant (see Appendix \ref{appendix:questions}). In response to each question, the discussion was permitted to flow naturally allowing the interviewer to enquire further about specific topics raised by the participant. For consistency, all interviews were performed by the same interviewer. A pilot interview was held with one participant to evaluate and improve the questionnaire used for the study.

Interviews were held remotely, using the Zoom video conferencing service provided by our institution. Zoom was configured to record a textual transcript of the interview which was saved after the interview was completed. Audio and video recordings were not stored. After each interview, to maintain participants' privacy, the transcript was de-identified such that the participant's name was replaced with a unique participant identifier (e.g., ``John Smith'' became ``P02'') and organizations were replaced with descriptive placeholders (e.g., ``Acme Corporation'' became ``EMPLOYER'').

\subsection{Data Analysis}

Data analysis involved several steps: 1) cleaning, 2) open coding, and 3) code analysis. These are discussed in turn below.

\subsubsection{Transcript Cleaning}\label{sec:method-cleaning}

After de-identification, the interview transcripts produced by the Zoom video conferencing service required cleaning prior to being used for open coding. The transcripts generated by Zoom were composed of lines of text, where each line was labelled with the current speaker. Two cleaning tasks were performed. First, filler or pause words (e.g., ``umm'', ``hmm'', etc.) were removed when they were the only text on a given line. These were inserted by Zoom when, as part of engaging in a natural conversation, the interviewer would provide positive feedback to the participant to indicate they understood the topic being described. Second, for spoken responses longer than a few seconds, Zoom would record multiple lines in the transcript for the same speaker, even if the response was all part of the same topic/discussion. To reduce the number of data items being coded (without adding any extra value), sequential transcript lines describing the same topic were combined.

\subsubsection{Open Coding}\label{sec:method-coding}

Open coding was used to analyze the interview transcripts. Two analysts independently reviewed each line in each (cleaned) transcript and assigned between one and three codes (i.e., labels) from a shared code book. All coding was performed by the same two individuals. Though analysts had access to the full transcript for context, only responses from the interviewees were coded. Code assignments did not have order or precedence.

In keeping with the open coding method, analysts were permitted to add new codes to the code book as new concepts were encountered or to extend the definition/description of a code.  The code book had two levels of hierarchy: a ``category'' and then ``detailed codes'', both of which were permitted to change as the analysis progressed. The code book was initially seeded with a handful of codes based on the authors' own industrial and research experiences.

After independently reviewing each transcript, the two analysts met to review the assigned codes and discuss their findings. In the meetings, each line of each transcript was reviewed for disagreements in the codes independently assigned by each analyst. If there was disagreement, then it was discussed in detail. After discussion, analysts could either change their assigned codes or leave them as they were (i.e., ``agree to disagree''). During these meetings, to reflect an improved understanding between the analysts, code definitions could be modified or new codes could be added to the code book. In many cases, the discussion between analysts was productive, leading to deeper insight about the concepts discussed by interviewees.

Two questions in the interview were multiple choice questions with a discrete set of answers (Questions \#10 and \#12). These questions were included to gather concrete data and also to stimulate additional discussion from the participant. The responses were still coded by both analysts (first independently, then meeting for discussion), but a pre-defined set of codes were used. While a short answer was accepted, in most cases participants provided both a short answer (their choice) and then a longer form justification, which was also subjected to open coding analysis. 

Agreement between the analysts was calculated at two points in time: 1) after independent analysis, and 2) after the analysts met to discuss their results. Since each analyst was required to assign at least one code, but could assign up to three codes, the agreement calculation needed to account for the possibility that analysts assigned both different codes and different numbers of codes. Agreement was calculated as follows: for each line in each transcript, compute the proportion of agreed codes to the total number of unique codes assigned for the line; then, calculate the average agreement over all lines in all transcripts. For instance, for a single line, if analyst \txt{R1} assigned codes \txt{A}, \txt{B}, \txt{C} and an \txt{R2} assigned codes \txt{B} and \txt{D}, then the agreement would be calculated as $1/4 = 0.25$ since there is one agreed code and four unique codes on the line.

\subsubsection{Analysis}

Analysis of the coded interview transcripts proceeded in three steps. First, a cleaning step was performed where transcript lines that both analysts agreed were ``unclear'' or contained unrelated discussion were removed.

Second, the codebook was reviewed for similar or duplicate codes and similar codes were combined with the necessary changes to their definitions.

Third, each transcript was collapsed into a ``bag of codes'' that described the participant's experiences and opinions related to AC development and confidence assessment. For a detailed code to be included in the ``bag'' for that participant, it must have satisfied all the following criteria: 

\begin{itemize}

    \item the code was assigned to at least one line in the transcript;

    \item both analysts (after the discussion phase) agreed that the code should be assigned to the transcript line; and

    \item the code is not already in the bag (to avoid duplicates).

\end{itemize}

Finally, trends across all $N$ bags of codes were analyzed, within the top-level categories defined by the codebook, to determine which concepts or themes were the most prominent across the study participants. Coded interview transcripts were also reviewed and quotes from participants were extracted to enrich the discussion of the data.

\section{Results}\label{sec:results}

This section presents the results from recruitment, participant demographics, and the open coding analysis.

\subsection{Recruitment and Participant Demographics}

During recruitment a total of 29 invitation emails were sent to prospective participants, from which 25 individuals responded indicating interest in participating in the study (response rate of $86\%$). Four of the respondents did not satisfy the inclusion criteria. The remaining 21 respondents were invited to schedule an interview, from which 19 interviews were conducted. One respondent did not reply to our request to schedule an interview and the other was unable to attend the agreed-upon interview time. During the interview, all 19 participants were (re-)confirmed to satisfy the inclusion criteria for the study. The final number of participants included in the study was $n=19$ (participation rate of $66\%$).

\paragraph{Experience.} Participants were asked to describe their number of years of professional work experience, both for general activities (not necessarily assurance related work) and working in systems assurance. Overall, the group of participants was relatively experienced, with an average of 23 years of general professional experience and 16 years of systems assurance experience. Additionally, seven participants had more than 35 years of general professional experience. Table \ref{tab:experience} contains a more detailed breakdown of the participants' experience.

\begin{table}
    \footnotesize
    \centering
    \def\arraystretch{1.25}
    \begin{tabular}{ccc}
        \toprule
        \multirow{2}{*}{\textbf{Years of Exp.}}
            & \multicolumn{2}{c}{\textbf{Num. Participants}} \\\cmidrule(lr){2-3}
            & General & ACs \\ \midrule

        0-5     & 1     & 4 \\ 
        6-10    & 4     & 4 \\ 
        11-20   & 5     & 4 \\ 
        21+     & 9     & 7 \\ \midrule
        \textbf{Avg.}    & \textbf{23}    & \textbf{16} \\\bottomrule

    \end{tabular}
    \caption{Duration of participant professional (``general'') experience and specific experience with ACs.}
    \label{tab:experience}
\end{table}

\paragraph{Industries.} Participants were asked to describe the industries they have worked within, both for general activities (not necessarily assurance related) and those for which they have contributed to an AC. The automotive industry (including autonomous vehicles) was the most well represented ($n=9$, $47\%$). Other well represented areas were aviation ($n=7$, $37\%$) and defence ($n=6$, $32\%$). Table \ref{tab:industies} contains a detailed breakdown of industries. Note that many participants reported experience in multiple industries.

\begin{table}
    \footnotesize
    \centering
    \def\arraystretch{1.25}
    \begin{tabular}{lcc}
        \toprule
        \multirow{2}{*}{\textbf{Industry}}
            & \multicolumn{2}{c}{\textbf{Num. Participants}} \\\cmidrule(lr){2-3}
            & General & ACs \\ \midrule

        Automotive (incl. autonomous vehicles)      & 9     & 9 \\ 
        Aviation                                    & 7     & 4 \\ 
        Defence                                     & 6     & 6 \\ 
        Health Systems / Medical Devices            & 4     & 3 \\ 
        Marine Systems                              & 2     & 0 \\ 
        Mining / Oil \& Gas                         & 4     & 4 \\
        Nuclear Systems                             & 3     & 3 \\
        Rail                                        & 6     & 4 \\
        Robotics and Automation                     & 3     & 2 \\ \bottomrule

    \end{tabular}
    \caption{Participant industries of experience, both general experience and specifically with ACs.}
    \label{tab:industies}
\end{table}

\paragraph{Roles.} Preparing an AC is a significant undertaking, typically performed by multiple persons who assume different roles. Participants were asked what role(s) they had assumed while contributing to ACs. The most common role was ``reviewer'' (i.e., peer or internal reviewer), with most ($n=16$, $84\%$) participants indicating they had experience reviewing ACs. Further, a majority of participants ($n=14$, $74\%$) also had experience contributing as a ``developer'' (i.e., author, creator). Less common roles included ``approver'' ($n=5$, $26\%$), ``manager'' ($n=4$, $21\%$), and ``auditor'' ($n=8$, $42\%$). Note that many participants reported experience with multiple roles.


\paragraph{Geography.} Though geography was not a key factor considered when sampling for this study, the sample we obtained contained participants from a range of locations: North America ($n=13$), the United Kingdom ($n=4$), and Europe ($n=2$).

\subsection{Open Coding}

After cleaning the interview transcripts as described in Section \ref{sec:method-cleaning} and filtering out questions and prompts from the interviewer, there were 685 transcript lines\footnote{We use the term ``line'' to refer to a sequence of related words, phrases, or sentences; each transcript line might take of many lines on a printed sheet of paper.} of spoken content from participants that were used for open coding. On average, each participant's transcript had 36 lines and the average length of transcript lines was 75.7 words. During coding 94 transcript lines were excluded from the analysis: 89 lines contained content that both analysts agreed was unrelated ``filler'' (e.g., off-topic discussion, greeting at the beginning of the interview, etc.); and 5 lines contained ``unclear'' discussion that the reviewers could not understand due to Zoom transcription errors or the participant's (verbal) response being jumbled. After excluding filler and unclear lines, the corpus for the study consisted of $n=589$ transcript lines.

\paragraph{Codebook.} An overview of the final codebook generated during open coding is shown in Table \ref{tab:codes}. Codes were organized into a two-level hierarchy with the first level consisting of six ``categories'' and the second level containing 54 detailed codes. Two additional categories (\it{Impact} and \it{Importance}) were used to code the responses of multiple-choice questions, which had a closed set of potential answers/responses for the participants to pick from. In total, 605 codes were assigned as part of the open coding analysis. The most common category of codes used was \it{Methods} (assigned to 207 lines) followed by \it{Barriers} (assigned to 135 lines), which is not surprising given that these topics are related to the core purpose of this study. Detailed results for each code category are presented below in Section \ref{sec:results-categories}.

\begin{table}
    \footnotesize
    \centering
    \def\arraystretch{1.25}
    \begin{tabular}{lcc}
        \toprule
        \bi{Category} & \bb{Num. Detailed Codes} & \bb{Num. Code Uses} \\ \midrule
        \bi{Motivation} for Preparing ACs                    & 10   & 69\\
        Quality \bi{Attributes} of ACs                       & 4    & 59\\
        \bi{Expression} of ACs                               & 7    & 60\\
        \bi{Methods} for Confidence Assessment               & 12   & 207\\
        \bi{Uses} of Confidence Assessment Results           & 7    & 75\\
        \bi{Barriers} to Conducting Confidence Assessment    & 14   & 135\\ \midrule
        \bi{Impact} of Confidence Assessment*                & 3*   & 28*\\ 
        \bi{Importance} of Confidence Assessment*            & 4*   & 20*\\ \midrule
        Total                                                & 61 (54) & 653 (605) \\\bottomrule
    \end{tabular}
    \caption{Overview of final codebook (* denotes a multiple-choice question).}
    \label{tab:codes}
\end{table}

\paragraph{Open Coding Agreement.} Each analyst independently coded each interview transcript before meeting to discuss (and potentially modify their assigned codes). Agreement between analysts was computed before and after their discussion at both the categorical and detailed code level, using the calculation described in Section \ref{sec:method-coding}. Across all interview transcripts, before discussion, the analysts agreed on an average of 71\% of code assignments per transcript line at the categorical level and an average of 57\% of assignments per transcript line at the detailed code level. After discussion, agreement was measured to be an average of 98\% per line at the categorical level and an average of 97\% per line at the detailed code level. For reference, for the final codebook consisting of 61 detailed codes, if codes were randomly assigned with equal probability, the average level of agreement per transcript line would be approximately $1.86\%$\footnote{We verified this using a statistical simulation of our agreement calculation over a 100,000 transcript lines that had randomly assigned codes, including accounting for the possibility that analysts picked different numbers of codes.}.

\subsection{Coded Results}\label{sec:results-categories}

This sub-section discusses the coded results for each top-level category in codebook.  Both categories and detailed codes are identified by \it{italics} style. Where possible, quotes from participants are provided to support and enrich the discussion. In cases where two or fewer participants discussed a topic, we have reported $\leq 2$ in the results. This was done protect the anonymity of participants, as the community of AC practitioners is relatively small and a participant's personal position on a topic might be well known.

\subsubsection{Motivations for Preparing ACs}\label{sec:results-motivation}

\begin{figure}
    \centering
    \includegraphics*[width=0.75\textwidth]{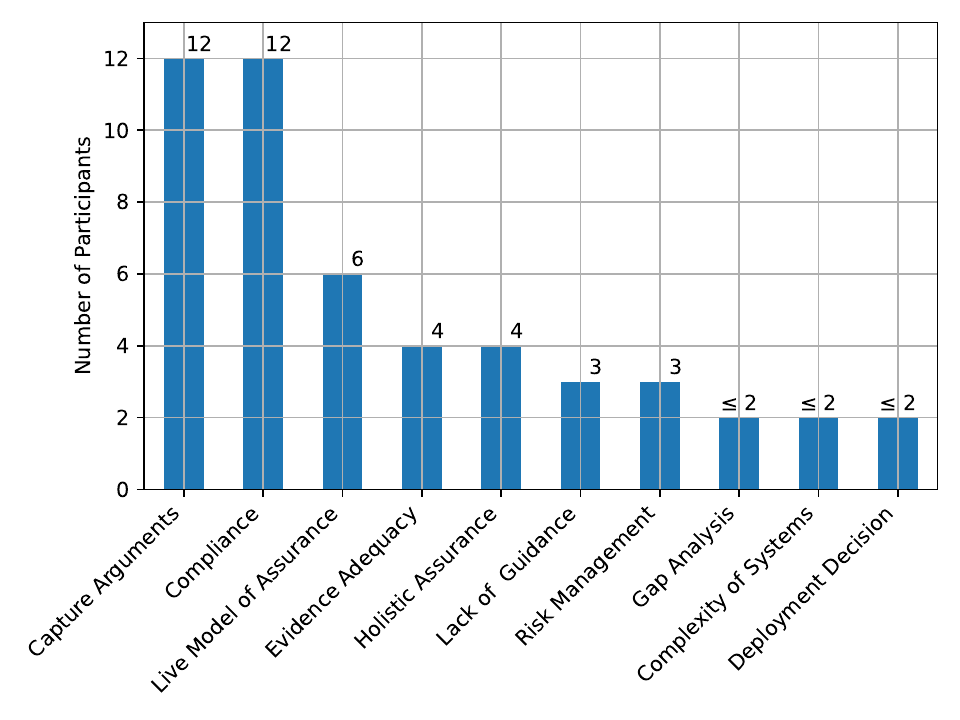}
    \caption{Number of participants who identified each \it{motivation} for preparing ACs.}
    \label{fig:motivation}
\end{figure}

The \it{Motivation} category of codes was used to capture the reason(s) that a participant (or their organization) might prepare an AC. Ten detailed codes within this category were identified which are shown alongside the number of participants that discussed each topic in Figure \ref{fig:motivation}. Most participants described multiple motivations for preparing an AC; on average, each was associated with $2.58$ detailed codes from the \it{motivation} category.

A common motivation for preparing an AC was to achieve \it{compliance} ($n=12$, 63\%) with technical standards, regulatory requirements, or contractual obligations. This is not in itself surprising given the importance that standards and regulations have in engineering for critical systems, see Section \ref{sec:introduction} for examples. ACs also have a history of use in some countries (e.g., the United Kingdom), where governments (or government-adjacent organizations) include them as deliverables in contracts. For instance, \Parti{P10} said: \textit{``Oh, for the main part it's [the AC] produced because there is a standard or a contractual obligation to produce the artifacts.''}. However, while compliance appears to be an initial factor motivating the creation of an AC, practitioners recognize other benefits that come from their use, as illustrated by \Parti{P02}: \it{``Initially, it [the motivation] was purely that the standard requires it. \ldots it was really just: it's a requirement of the standard that we were working to at the time and therefore we should put one together. Since then, of course, I have realized that it's [preparing an AC] much more important than that.''}

The other most common motivation for preparing an AC was to \it{capture} [assurance] \it{arguments} ($n=12$, 63\%) in a structured or systematic manner. Indeed, organizing assurance-related information has been a core value proposition of structured argumentation approaches from their inception \cite{kelly}. Nonetheless, it is useful to recognize that this continues to be an important motivator for preparing ACs, as illustrated by \Parti{P06} when they described how an AC helped them organize a large body of safety-related documentation: 

\begin{quote}
\it{
``\ldots there was lots of extensive documentation about, various safety aspects of the system but then they weren't particularly easy to navigate by themselves, so the assurance case effectively served as a way to quickly structure a lot of that information \ldots where the role that each piece of evidence instead of just, you know, being spread out over hundreds of pages of documents, but so that was logically connected together to an argument'' 
}
\end{quote}

Another notable motivator for preparing ACs that participants identified was to build a \it{``live'' model of assurance} ($n=6$, 32\%) that can be updated regularly as the system evolves, both during development and after deployment. This notion was well described by \Parti{P14}: \it{``\ldots it's not like you, you know, you put a bow on the assurance case and say, `we're done'. You know, you're constantly having to go back and revisit it. So, it's like continuous process. It's a live [artifact]''}. And also by \Parti{P10}: \it{``\ldots one of the objectives that we have is to try to get people to see it as a live activity rather than a retrospective [activity].''}

\subsubsection{Expression of ACs}

The \it{Expression} category considered notations or methods used by practitioners to represent or document ACs. These can be broadly distinguished between structured notations (e.g., GSN) and narrative (e.g., written safety reports). Many participants described experience with both structured notations and narratives ($n=11$, 58\%), though not necessarily combining them.

A strong majority ($n=17$, 0.89\%) reported experience with structured notation, with a minority of participants ($n=6$, 32\%) indicating that they were the only method they had used to express ACs. Many participants had positive views on structured notations, indicating that they helped them and their colleagues organize assurance information (e.g., see the quote from \Parti{P06} above). However, some ($n=4$, 21\%) participants expressed negative opinions. For instance, \Parti{P11} raised concerns about syntax and semantic details distracting from the essential aspects of assurance arguments: \it{``\ldots the main thing is that there's all kinds of stuff added that's unnecessary. Arguments are about premises and reasonings. They're not about context. They're not about solutions. They're not about evidence. All of those, they're not about packages.''}.

Narratives have been used by a majority of participants ($n=13$, 68\%) with a small number ($n=2$, 11\%) having used exclusively narrative expression.

\begin{figure}
    \centering
    \includegraphics*[width=0.80\textwidth]{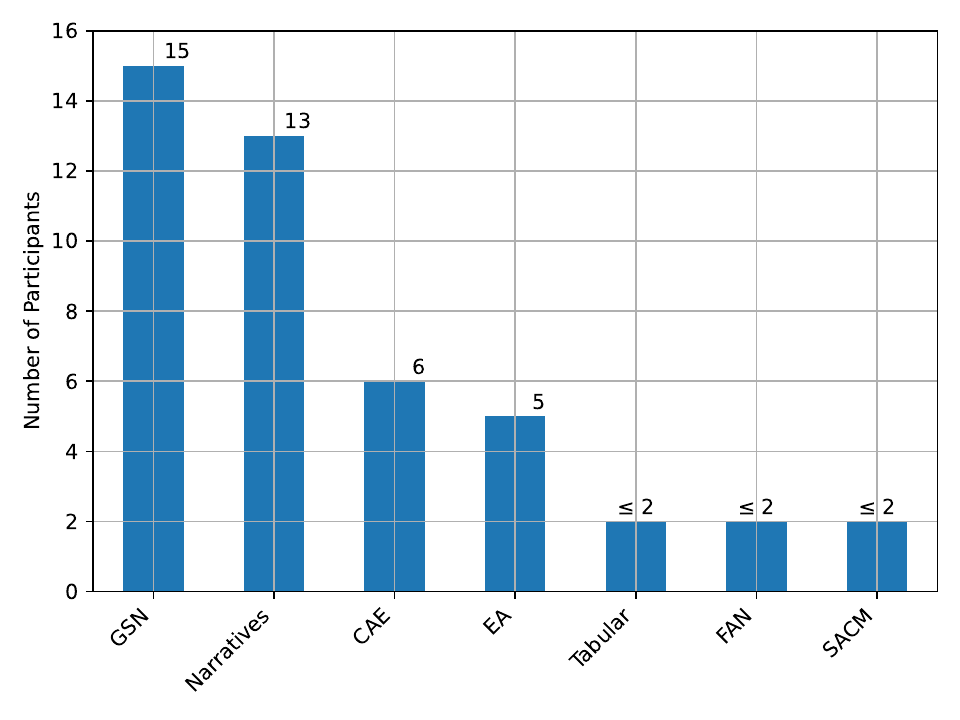}
    \caption{Number of participants who identified each method \it{expression} for ACs.}
    \label{fig:express}
\end{figure}

A breakdown of the detailed codes for this category is provided in Figure \ref{fig:express}. All participants reported at least one method of expression, with an average of 2.26 codes applied per participant.

\subsubsection{Quality Attributes for ACs}\label{sec:results-quality-attributes}

Fundamentally, ACs are intended to capture an argument about one or more quality attributes of a system. The interview included a question which asked participants to describe the quality attributes that were argued in the ACs they have worked with. Unsurprisingly, a strong majority ($n=16$, 84\%) of participants indicated they had prepared an AC where \it{safety} was the main quality attribute being considered. Only four participants ($n=4$, 22\%) had prepared an AC where \it{security} was the primary quality attribute. However, many participants ($n=11$, 58\%) had experience preparing ACs that \it{mixed many quality attributes}, including safety, security, reliability, availability, fitness-for-purpose, and compliance. Finally, two participants ($n=2$, 11\%) indicated they had worked with at least one AC that \it{mixed safety and security}, but one of these participants described a negative experience in this regard.

\subsubsection{Methods for Assessing Confidence in ACs}\label{sec:results-methods}

\begin{figure}
    \centering
    \includegraphics*[width=0.75\textwidth]{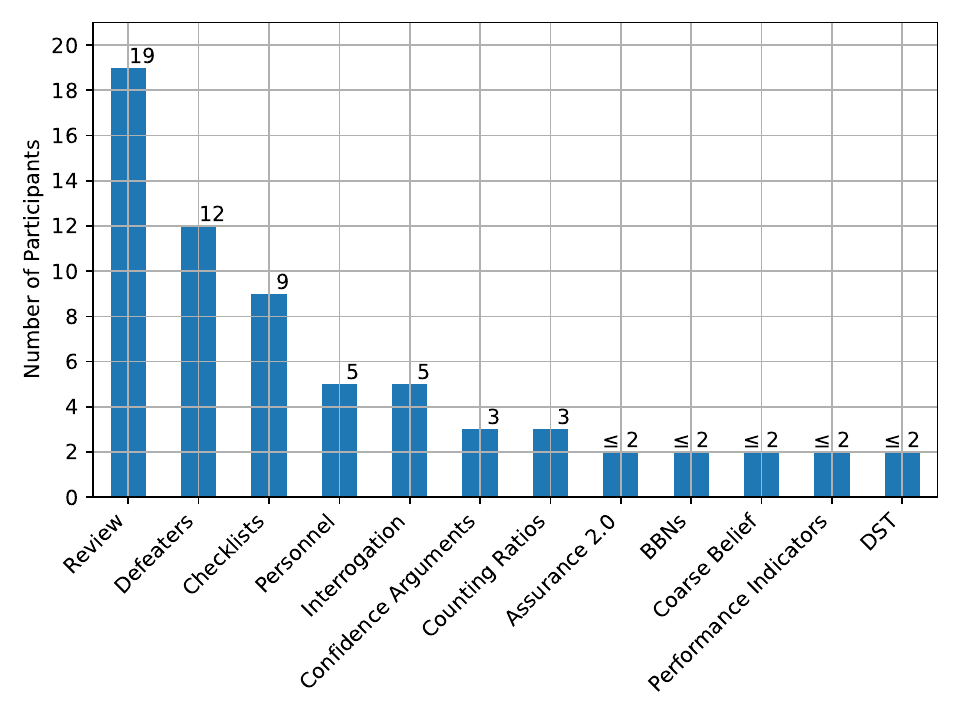}
    \caption{Number of participants who identified each \it{method} for gaining confidence in ACs.}
    \label{fig:methods}
\end{figure}

For the open coding analysis, we adopted a broad definition of ``method'' such that the \it{Methods} category considered methods, techniques, activities, practices, or procedures that have been used by practitioners to assess or increase confidence in an AC. For example, conducting a peer-review activity of an AC was coded in this category, even if a specific review methodology with prescribed steps was not described by the participant. Figure \ref{fig:methods} shows a breakdown of the methods used by participants. All participants reported at least one CAM, with an average of 3.47 method-related codes applied per participant. More details are provided for the most popular methods below.

\paragraph{Method: Review.}

Every participant ($n=19$, 100\%) indicated they had used some form of independent or peer review as a CAM. While this result is not itself surprising, since review is a cornerstone of engineering practice, it is still notable that all participants identified this method, with many participants describing review (and the corresponding dialogue amongst reviewers and authors) as an essential activity for increasing confidence in an AC. For example, \Parti{P15} said \it{``having, you know, a big open discussion between several reviewers and the authors of the case is, for me, is the best way to [gain] qualitative confidence.''} However, some participants also described concerns with, or limitations of, review. For instance, \Parti{P16} observed that review is prioritized more than it should be: \it{``\ldots there are a lot of things that we currently do, but we put a lot of store in peer review. Probably more store than I'm comfortable with people putting on peer review.''}

Some participants ($n=5$, 26\%) described a specific type of review where the reviewer systematically examines each reasoning ``step'' in the argument and considers whether the premises (child claim(s)) support the conclusion (parent claim). We have coded these as a separate activity called ``interrogation''. When participants described an interrogation method, it was usually contrasted with a less focused type of review such as checking for spelling/grammar issues or checking the notation or style of an argument, without considering the soundness of the argument itself.

\paragraph{Method: Defeaters.}

A majority of participants ($n=12$, 63\%) described experiences using dialectic arguments to challenge or defeat positive arguments in an AC. We have collectively coded these as ``defeaters''. Defeaters are recognized by many authors (from both academic and industrial backgrounds) as a method of mitigating confirmation bias in an argument (see Section \ref{sec:background}), so it is not surprising that they are well represented in the coded results. \Parti{P02} described the impact that using defeaters had on their organization's ACs: 

\begin{quote}
\it{``\ldots this was the big eye opener. We had created a number of safety cases for each release of our product \ldots and for the first time, we used `eliminative induction', `doubting', or whatever. And it was remarkable. We discovered something like 25 problems that we had not seen previously. Although we had been producing safety cases. We have not seen these particular problems. Some of those problems we could immediately fix \ldots something like 12 or so of them were problems that could not be easily fixed.''}
\end{quote}

While participants generally described their experience with defeaters as valuable, they identified challenges with their application:

\begin{itemize}
    
    \item Since defeater identification is an open-ended activity, it is not always clear when to stop identifying new defeaters. That is, there are no clear ``stopping rules''. This is in contrast with other methods that practitioners often use (e.g., Failure Modes and Effects Analysis (FMEA)) for which there are criteria or guidelines that help to determine when the analysis is complete.

    \item Once resolved or rebutted, it is not clear whether defeaters should be left in the AC's argument structure or removed. On one hand, defeaters can capture genuine problems and their resolution, which some practitioners viewed as contributing to increased confidence. However, some noted that leaving defeaters in the final version of the argument (even if they are marked as resolved) might invite undue criticism by those not familiar with dialectic techniques.

    \item In practice, it can be difficult to determine whether a concept should be framed as a defeater or as a positive claim. Indeed, it is often possible to adjust the wording (and notation) of a defeater to make it a claim, and vice versa. This can lead to ACs with many defeaters that could have been claims, which might artificially increase confidence in the case.
    
\end{itemize}

\paragraph{Method: Checklists.}

Just under half of participants ($n=9$, 47\%) indicated they had used ``checklists'', generated from guidance documents or standards, as a means of assessing confidence in the case. In many cases, internationally recognized standards (e.g., UL 4600 for autonomous vehicles \cite{ul4600}) were used as the basis for the checklist, and in other cases internal organization-specific checklists were used. While compliance audits/assessments are a form of checklist-based assessment, they were not the primary use of checklists described by participants. More often participants described these activities as ``gap analyses'' or simply checking that the topics were addressed in the case. For instance, \Parti{P12} described their use of checklists as: \it{``Reviewers review against different checklists. Checklist that we built up based on our experience, but also like things from the literature''} and \Parti{P08} said \it{``\ldots [we perform] reviews and then also reviewing documents [checklists] to make sure they key topics in each area are covered.''}

\paragraph{Method: Quantitative Assessments.}

Participants were asked about their knowledge of, experience applying, and opinion of quantitative confidence assessment methods (see Section \ref{sec:background}). Table \ref{tab:quantitative} shows both the level of awareness and opinion of quantitative confidence assessment methods reported by participants.

\begin{table}
    \footnotesize
    \centering
    \def\arraystretch{1.25}
    \begin{tabular}{clcccc|c}
        \toprule
            &&\multicolumn{5}{c}{\bb{Opinion of Quantitative Methods}} \\\cmidrule(lr){3-7}
            &&Negative & Positive & No Opinion & Unclear & \bb{Total}\\ \midrule
            \multirow{5}{2cm}{\bb{Awareness of Quant. Methods}}
                & Not Aware         & 1 & 0 & 2*& 0 & 3 \\
                & Aware, Not Used   & 11& 0 & 0 & 2 & 13\\
                & Aware, Used       & 0 & 1 & 0 & 1 & 2 \\
                & Unclear           & 1 & 0 & 0 & 0 & 1 \\  \cmidrule(lr){2-7}
                & \bb{Total}        & 13& 1 & 2 & 3 & 19 \\
        \bottomrule
    \end{tabular}
    \caption{Awareness and opinion of quantitative confidence assessments (* denotes ``no opinion because no awareness'').}
    \label{tab:quantitative}
\end{table}

Most participants ($n=13$, 68\%) indicated they were aware of quantitative CAMs but had not applied them to real-world systems (though some had tried applying them to small ``toy'' examples as a learning exercise). In fact, only two participants ($n=2$, 11\%) reported using a quantitative method for a real-world system. The remainder of participants were either not aware of any quantitative methods or provided unclear responses that could not be coded.

In terms of opinion, a majority ($n=13$, 68\%) of participants had a negative opinion of quantitative CAMs. One participant was not aware of any quantitative methods before the interview, but then expressed a negative opinion about the notion of quantitative assessments once the concept was explained to them by the interviewer. Reasons for negative opinions included: 

\begin{itemize}

    \item A lack of confidence that the methods themselves produce trustworthy results. For instance, \Parti{P01} said \it{``\ldots you have quantitative methods and don't really know if they work or not. So, it's a huge problem for using it.''}. Some participants went further and pointed out that \it{``a number with no basis in reality is worse than no number''}. That is, since quantitative methods are (rightly) trusted in many other areas of engineering and science, interest holders without extensive knowledge of ACs and CAMs might mistakenly place their trust in an unreliable number, whereas if no number exists then they are required to consider alternative means of evaluating confidence.

    \item Concerns that ACs contain more nuanced information, which cannot be easily distilled down into a handful of numbers. For example, \Parti{P10} said \it{``\ldots it seems to me that it's consolidating a lot of very uncertain information and producing a number, and the amount of confidence that you can have in the number I don't think warrants the amount of effort that goes into creating [it].''}

    \item It might be difficult to justify a specific confidence value to an independent assessor or regulator, especially if they are not familiar with the specific method(s) being used. \Parti{P14} said \it{``\ldots if we ever had to go and argue this in front of [a regulator], \ldots with 86\% confidence that the system wouldn't [harm someone]. Well, then the next question would be, how did you prove that? \ldots We would have to start going down some mathematical rabbit holes where, you know, it may become more difficult to defend that kind of thing \ldots''}

    \item There are challenges identifying the inputs to quantitative methods, i.e., the numbers assigned at the leaves of the argument tree. Existing methods are, perhaps rightly, sensitive to their leaf inputs. But distilling the qualitative information in the argument's leaves into numerical values might be difficult to do reliably. For instance, \Parti{P16} said \it{``You know, people might, be very good at understanding the processes that they're using, but actually making it into a [number]? I don't really have confidence that people can do it. And, putting all those [numbers] into a big mathematical equation just doesn't fill me with a great deal of confidence \ldots''}

\end{itemize}

Only \Parti{P02} reported a positive opinion of quantitative confidence assessment: \it{``The Bayesian [method] it seems, is much easier for experts to give you an assessment of a post-condition on something, and to give a prior and so taking an arbitrary prior and taking the experts decisions on the post-conditions is, I have found to be useful.''} In other words, to determine the final assessment (the ``post-condition'') asking analysts to react to a potentially incorrect (the ``prior'') assessment of confidence produces more useful assessments.

\subsubsection{Uses of Confidence Assessment}

Participants were asked to describe how they use the results of CAMs. The open coding analysis identified seven detailed codes within \it{Uses} category. All participants reported at least one use of confidence assessment results, with an average of 2.84 detailed codes per participant. 

%

A majority ($n=14$, 73\%) indicated that they have used the results of confidence assessment to communicate with interest holders. For instance, the results of a review or checklist activity might be shown to an organization's decision makers as a way of communicating the maturity of the AC. This is related to the goal of understanding (qualitative) risk associated with the system, which participants ($n=8$, 42\%) also identified as a use of confidence assessment. A minority of participants indicated they had used confidence assessment results to evaluate the argument ($n=7$, 37\%) and supporting evidence ($n=6$, 32\%) and to address issues ($n=8$, 42\%) to improve the AC's quality. Avoiding confirmation bias was also identified by some participants ($n=5$, 26\%). Finally, it is notable that only a minority of participants ($n=6$, 32\%) described using confidence assessment results to support a decision to deploy a system. 

\subsubsection{Barriers to Performing Confidence Assessment}\label{sec:results-barriers}

\begin{figure}
    \centering
    \includegraphics*[width=0.75\textwidth]{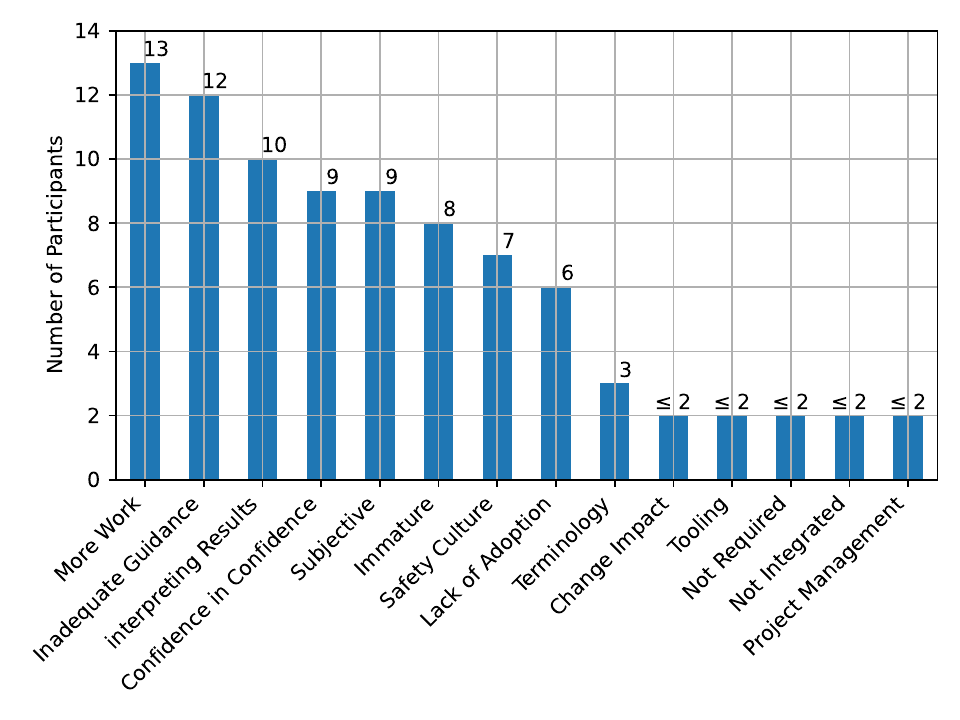}
    \caption{Number of participants who identified each \it{barrier} to performing AC confidence assessment(s).}
    \label{fig:barriers}
\end{figure}

Participants were asked to discuss reasons that a CAM might not be used (i.e., ``barriers''). For the purpose of this work, the CAM could be a method the participants had used (see the \it{Methods} category in Section \ref{sec:results-methods} above) or an alternative activity that the participant had not used, but thought would be difficult to apply in practice. The open coding analysis identified 14 detailed codes in the \it{Barriers} category. All participants identified at least one barrier, with an average of 4.47 detailed codes assigned per participant. Figure \ref{fig:barriers} shows the number of participants who identified each barrier. Several of the barriers are discussed in more detail below.

\paragraph{Barrier: More Work.} A majority ($n=13$, 68\%) of participants identified that additional effort is a barrier to using a CAM. In reference to the argument interrogation method, \Parti{P16} said that \it{``\ldots if you are writing a healthy GSN safety case and you ask people to look at every decomposition, that can start to get quite labour intensive.''} Further, several participants observed that it might be difficult to justify the extra effort to managers responsible for project schedules and budgets. For instance, in the context of a quantitative assessment method, \Parti{P02} said \it{``You're asking me basically to delay the project by looking at these confidence levels \ldots There is a managerial buy-in required to make that a positive thing. There's a lot of work there and there's no, formal [widely accepted] method of doing it.''} 

Some participants noted that some quantitative CAMs require many numerical inputs that an analyst needs to select. \Parti{P16} said \it{``[The] level of effort to use methods could be high. Especially if [you are] needing to put many numbers all around.''} Indeed, the DST method from \cite{idmessaoud2024} requires four numerical inputs per leaf node (two for the node, two for the connection to the parent) and an additional three inputs per parent node. Similarly, the BBN method from \cite{hobbs2012, diemert2024} requires two inputs per leaf (one for node, one for connection to parent) and another two inputs per parent node. In either case, for an industrial scale AC such as the one in \cite{millet2023}, which contains on the order of 500 nodes, it is a significant task to select (or even review) all of these assignments.

\paragraph{Barrier: Inadequate Guidance.} Many participants ($n=12$, 63\%) noted that even though (qualitative or quantitative) CAMs exist, they felt that the methods are inaccessible to practitioners due to inadequate published guidance or training materials about how to apply these methods in practice. This detailed code was used to capture a range of concerns that participants had, including: 

\begin{itemize}

    \item In order to use a method, practitioners need to understand the theory supporting the method so they can justify it to their interest holders. The description of some method(s) in the literature can be difficult for practitioners to understand.  This is especially true for quantitative methods that are expressed with mathematical notation(s) some of which are based in theoretical frameworks that are not familiar to practitioners. This problem was clearly articulated by \Parti{P01}: \it{``I think another problem is that you don't necessarily understand how the confidence algorithm is working''} and again by \Parti{P16}: \it{``\ldots industrial practice is a long way from, academic practice around safety cases. So, the research that's kinda like the leading-edge research is so far, removed from what practices people need to get up a really big learning curve before you can actually apply it.''}

    \item Since there are multiple methods reported in the literature, it can be difficult to know which method to use. Again, \Parti{P01} described this well: \it{``Like, you have multiple methods. I would like to know which one to use when. They obviously don't work for everything, there might be some that are really good for one thing and not another. I don't know which one.''}
    
    \item \Parti{P19} noted that ACs and their assessment are not typically covered as part of the standard system or software engineering courses offered by educational institutes. This suggests there will continue to be a knowledge gap in this area for new graduates from training programs.

    \item Technical standards and regulations, which prescribe the creation of ACs, do not provide enough guidance on how to construct an AC or evaluate confidence in the argument, which was described by \Parti{P18}: \it{``I want to blame a little bit [technical] standards, because standards they actually say to [prepare an AC], and they didn't tell how you can [assess confidence].''} Since complying with standards and regulations is often a requirement for practitioners, providing clearer guidance on how to perform such an assessment would benefit practitioners.

\end{itemize}

\paragraph{Barriers: Interpreting Results and Subjectivity.} Just over half of the participants ($n=10$, 52\%) identified barriers related to difficulty interpreting the results from a quantitative confidence assessment. For example, if the BBN confidence assessment produces subjective probability $0.82$, how should this be interpreted? Is $0.82$ high confidence? Is it enough to deploy a system, or is more work required? What if, after more effort, the value were to increase to $0.87$, is that sufficient? Several participants identified a related concern about the subjectivity  of confidence assessment ($n=9$, 47\%), where different practitioners arrive at different results for either a qualitative or quantitative assessment. Indeed, system safety practitioners will all have unique experiences and industry-specific knowledge that might cause them to weight various arguments or evidence differently. In the context of quantitative assessments, \Parti{P12} expressed this concern well: \it{``it could be that I put some numbers, and you put totally different numbers, and then who do we trust?''}

\paragraph{Barrier: Immature Methods.} Some participants ($n=8$, 42\%) expressed concerns about using immature (quantitative) assessment methods. For instance, \Parti{P01} said \it{``\ldots I think you have quantitative methods and don't really know if they work or not. So, it s a huge problem for using [them].''}

\paragraph{Barrier: Confidence in Confidence.} Several participants ($n=9$, 47\%) described a meta problem with confidence about how to become confident that the CAM produced a trustworthy result, which was captured with the detailed code \it{confidence in confidence}. In this regard participants expressed concerns with both qualitative and quantitative CAMs. For quantitative CAMs, concerns were focused on interpretation of results and subjectivity or the maturity of methods (see above). For qualitative methods, participants raised concerns about completeness of the assessment. For instance, if defeaters are used, it is not clear when to stop identifying defeaters to the argument, which was described by \Parti{P10}: \it{``[I have] started to [use defeaters], and I do find it useful. I think that there are challenges with recording it and and with the stopping rule \ldots''}.

\paragraph{Barriers: AC Adoption and Safety Culture.} Some participants ($n=6$, 32\%) suggested that the general adoption of ACs remains a challenge. That is, it is not possible to apply a CAM without first preparing an AC to assess. Further, participants ($n=7$, 37\%) identified issues related to safety culture as a barrier to performing confidence assessments. In short, the results of a confidence assessment might be unsavory, since performing confidence assessment requires additional effort and might produce results that require even more effort to address, so it is avoided. For instance, \Parti{P12} said \it{``\ldots that does happen in some sort of industries \ldots the safety case or safety analysis or even the safety culture is seen as a hindrance rather than a benefit.''}

\subsubsection{Impact and Importance of Confidence Assessment}

As part of the interview, participants were asked two multiple choice questions. These were asked to capture concrete (short answer) responses and also to elicit further discussion, which was used as part of the open coding analysis.

The first question was about the impact that CAMs had on the overall project. Participants were asked to answer based on their practical experience rather than describing an idealized outcome. Overall, participants strongly indicated that confidence assessment activities had a positive ($n=17$, 89\%) impact on their past projects. Only one participant ($n=1$, 5\%) indicated it had a neutral impact. Three ($n=3$, 16\%) participants had experiences where confidence assessment had a negative impact. These negative experiences were limited to specific projects, and the same participants also described other projects where confidence assessment had a positive impact. One participant did not clearly respond to this question.

The second question asked the participants for their general opinion, independently of specific project experiences they might have had, about the importance of CAMs and the extent to which they should be prescribed in the future. Almost all participants felt that confidence assessment should be either \it{mandatory} ($n=8$, 42\%) or at least \it{recommended} ($n=8$, 42\%). Only one participant ($n=1$, 5\%) thought it should be \it{optional} and none thought it should be a \it{discouraged} practice. One participant did not clearly answer this question. Many participants who answered \it{recommended} or \it{optional} indicated that they would have picked \it{mandatory} but for the fact that there are no minimum criteria or widely recognized methods (except for ``review'') and so it would be unreasonable to demand practitioners perform this activity without there being widely recognized and trusted methods for performing it.

\section{Comparison with Previous Studies}\label{sec:related}

\newcommand{\TandO}{T{\"o}rner and {\"O}hman}

Overall, there are relatively few empirical studies on ACs. Real-world ACs often contain confidential information (e.g., intellectual property, residual risks) that organizations do not wish to be made public. Therefore, it is difficult for researchers to obtain full ACs to study. There are some exceptions, such as \cite{rees2023}, but these are rare. Therefore, most empirical work (including our present paper) has focused on collecting data from practitioners using interviews and questionnaires about their use of ACs and related methods. While other researchers have engaged practitioners about specific topics related to ACs, to our knowledge, the present paper is the first empirical work on practitioner's use of CAMs. This section reviews other empirical studies that collected data from practitioners about their use of ACs and compares their results with those presented in this paper, to the extent that there is overlap; a summary is provided in Table \ref{tab:previous}.

\begin{table}[h]
    \newcommand{\lightrule}{\arrayrulecolor{gray}\midrule\arrayrulecolor{black}}
    \def\arraystretch{1.5}
    \scriptsize
    \centering
    \begin{tabular}{p{1.6cm}clcp{4cm}p{5.5cm}}
        \toprule
        \bb{Study} 
            & \bb{Year}
            & \bb{Method}
            & \bb{$n$}
            & \bb{Focus}
            & \bb{Relevant Result(s)}\\
        \midrule
        \TandO{} \cite{torner2008}
            & 2008
            & Interviews
            & 10
            & Adoption and uses of safety cases by Swedish automotive manufacturers.
            & ACs are useful as communication aids and for documenting arguments, but practitioners were concerned about the added work required to prepare ACs.
        \\
        Nair \etal{} \cite{nair2015}
            & 2012
            & Questionnaire
            & 52
            & Evidence management for critical systems and comparison between practices in industry v. academic literature.
            & Textual methods are used for expressing ACs, but there is interest in structured notations (e.g., GSN). Assessing confidence is important, but remains challenging for practitioners.
        \\
        Doss and Kelly \cite{doss2016}
            & 2016
            & Questionnaire
            & 31
            & Relationship between safety engineering activities (incl. ACs) and agile process approaches.
            & Iteratively developing and (re-)evaluating an AC is an important task for safety-critical systems.
        \\
        Cheng \etal{} \cite{cheng2018}
            & 2018
            & Interviews
            & 9
            & Understanding practitioner perception(s) of ACs.
            & ACs are useful as a communication aid. There are challenges related to scalability, practitioner expertise, confidence management, and tool support.
        \\
        Almendra \etal{} \cite{almendra2022}
            & 2022
            & Questionnaire
            & 31
            & Interplay between requirements and ACs.
            & Creation of ACs is driven by regulatory or organizational requirements. Structured methods for expressing ACs (e.g., GSN) are used by many practitioners.
            \\
        \lightrule
        \it{(this study)}
            & 2024
            & Interviews
            & 19
            & Methods and barriers for AC confidence assessments performed by practitioners.
            & N/A - \it{This is the present study}.
            \\
        \bottomrule
    \end{tabular}
    \caption{Summary of previous empirical studies on AC usage. Present study shown for reference.}
    \label{tab:previous}
\end{table}

\paragraph{\TandO{}.} In 2008, \TandO{} used interviews to survey ten engineers from three Swedish automotive companies about the adoption drivers, potential use cases, and issues for safety cases \cite{torner2008}. Overall, they concluded that there is a need for ACs for automotive technology. This survey was conducted before 2011, when the first edition of ISO 26262 was published, which required the preparation of an AC for electronics and software in automotive systems. Therefore, the survey's results provide some indication of the automotive industry's perspective on ACs prior to their inclusion in a landmark technical standard. The survey found that improvements in systems development (e.g., organizing technical information), project management (e.g., facilitation of technical tasks), and safety management (e.g., determining what constitutes ``acceptable'' risk) were thought to be drivers to adoption of ACs. Key use cases identified by participants included communication, documentation, and aiding early system development. Potential issues related to the use of ACs were identified. Participants expected that preparing an AC would result in an ``increase in workload'' since ACs do not replace another activity. Participants also worried about access to information required to prepare an AC, especially from suppliers that might not share technical details to protect intellectual property. Finally, participants worried about legal aspects related to product liability, regulation, and competence of practitioners to prepare ACs.

Even after 16+ years, some of \TandO{}'s findings are consistent with those from our work in several ways. First, the use of an AC (or in our case, confidence assessment) as a communication tool continues to be an important theme. Second, the motivation of preparing an AC to document assurance arguments is a motivation that we also identified (see Section \ref{sec:results-motivation}).  Finally, the concerns about ``increase in workload'' associated with preparing an AC (or performing confidence assessment) is another similarity. However, the motivations reported by our participants differ from some drivers of adoption described by \TandO{}. It is possible that the motivations for preparing ACs have shifted since 2008. For example, now that more industrial standards require the preparation of an AC than in 2008, achieving compliance is naturally a priority for practitioners.

\paragraph{Nair \etal{}} In 2012 Nair \etal{} used a web-based questionnaire to survey 52 practitioners to gain insight into how engineering evidence is managed for safety-critical systems in practice (rather than in academic literature) \cite{nair2015}. Their survey included participants from a range of industries, with strong representation from the aerospace, rail, automotive, and defence industries. 

Nair \etal{}'s results show that, at the time of their survey, graphical and structured methods (e.g., GSN, CAE, etc.) for representing ACs and organizing evidence were rarely used by practitioners who prefer textual methods (e.g., structured narratives or templates). However, there appeared to be interest in these methods since ``how to effectively create and structure safety cases'' \cite{nair2015} was identified as an important challenge for practitioners. Practitioners also felt that using argumentation to connect evidence to requirements from technical standards was important. Twelve years later, the results from our study suggest there has been increased adoption of structured argumentation methods: a strong majority of participants in our study ($n=17$, 89\%) reported experience with structured notations for expressing arguments. Though, it is worth noting our results show that narrative approaches are still used by practitioners as well.

Regarding confidence assessment, Nair \etal{}'s results also show that evaluating how well evidence supports a claim or objective from a technical standard was an important or very important challenge. Moreover, assessing confidence was found to be of high importance for practitioners, which is consistent with our result showing that confidence assessing activities generally had a positive impact on a project and that participants thought they were important to apply.

\paragraph{Doss and Kelly.} In 2016 Doss and Kelly used a questionnaire to survey 31 practitioners aimed at understanding challenges and opportunities for applying agile development methods to safety-critical systems \cite{doss2016}. A sub-set of their survey questions focused on AC development. Overall, the surveyed practitioners had a positive view of agile methods for safety-critical system development. Additionally, most practitioners recognized the importance of regularly (re-)evaluating the AC for a safety-critical system. Based on the survey results, Doss and Kelly proposed that incrementally developing an AC as part of an agile approach would be beneficial.

A minority of participants ($n=6$, 32\%) in our study identified that having a ``live'' model of assurance was a motivation for preparing assurance cases. This is much lower than Doss and Kelly's reported 90\% of participants.  Indeed, the purpose of Doss and Kelly's work survey was to investigate the use of agile methods, whereas our present study did not provide the same conceptual framing for the question, so it is not surprising that Doss and Kelly found this to be a stronger theme. Nonetheless, it is still a shared theme between the two works.

\paragraph{Cheng \etal{}} In 2018 Cheng \etal{} interviewed nine experts with experience developing safety-critical systems and preparing ACs \cite{cheng2018}. Their interviews focused on practitioner perceptions' of ACs. Benefits of ACs identified by Cheng \etal{} included: 1) ACs' value as an easy-to-understand communication aid; 2) ACs' use to organize thinking about safety in a project; and 3) ACs' applicability to novel technologies where standard approaches do not yet exist. Cheng \etal{}'s finding that ACs are a communication aid is consistent with our own results: a majority of participants ($n=14$, 73\%) reported using the results of AC confidence assessment to communicate with interest holders.

Participants identified several challenges associated with using ACs, including scalability, change management, practitioner expertise and skills, dealing with complex systems, managing uncertainty and confidence, the method being ``too flexible'', and handling incomplete information. Of these, scalability and change management were rated as the most significant challenges. The absence of tool support to address these challenges was noted as a compounding factor that made the challenges more acute. While the focus of Cheng \etal{}'s and our studies are different, there are notable similarities in terms of challenges reported by practitioners. First, Cheng \etal{} identified managing uncertainty and confidence as a challenge for practitioners, which is the focus of our present study, and our results suggest that this is far from a settled question from the perspective of practitioners. Second, the issue of scalability appears in our results in terms of the number of inputs required to use quantitative CAMs. Third, Cheng \etal{}'s findings regarding practitioner expertise and skills parallels our own results about inadequate guidance being available to practitioners. Finally, while Cheng \etal{} found tool support to be a challenge, and this was reported by a few participants, it was not a significant theme in our results.

\paragraph{Almendra \etal{}} Almendra \etal{} used a web-based questionnaire to survey a population of 31 practitioners and researchers about the interplay between requirements engineering and ACs \cite{almendra2022}. As part of their questionnaire, to contextualize their work on ACs and requirements, they also asked questions about the general motivations and use of ACs. Their results show that AC development is thought to be beneficial, but that most AC development is ultimately driven by regulatory requirements. For instance, a majority (at least 84\%) of their participants indicated that an important motivation for preparing ACs is to comply with requirements or recommendations from regulators or their own organization. Their finding that compliance is a significant motivating factor for AC preparation is consistent with results of our present study.

In terms of expression of ACs, Almendra \etal{} reported that a slight majority of their participants used GSN to structure ACs (52\%), with a smaller proportion (26\%) using textual narratives to express ACs. The results of our present study are partially consistent with Almendra \etal{}'s. While a majority of participants in our study also reported use of structured notations for expressing ACs, a notable number also described experiences with narrative or textual expression, which differs from Almendra \etal{}'s finding.

Overall, Almendra \etal{} found that there is a perceived benefit to developing system requirements and ACs concurrently. Challenges identified by survey participants included a lack of tool support, automation, and integration of tools with the larger system development lifecycle. Interestingly, in our study a lack of tool support was not a major theme identified by our open coding analysis. However, this difference might be due to the differences in objectives between Almendra \etal{}'s study and our own. Lack of tool support might be a concern for integrating ACs and requirements, but it might be less of a concern for confidence assessment. Alternatively, during our interviews, participants might have identified other barriers to confidence assessment which they felt were more important (e.g., inadequate guidance on confidence assessment), rather than focusing on a lack of tool support.

\section{Discussion}\label{sec:discussion}

The objective of this study was to develop an understanding of current practices for increasing or assessing confidence in ACs. The following discussion describes the current state of AC and confidence assessment practice within our study population. Then it presents considerations for researchers working on CAMs. Finally, this section concludes with a discussion of our study's limitations and threats to validity.

\subsection{Current State of Practice}

This sub-section summarizes the current state of practice for our study population in terms of the categories identified in our study's open coding analysis.

\paragraph{Motivations for Preparing ACs.} Compliance with regulations, technical standards, contractual, or organizational requirements is a principal motivator for preparing ACs. Practitioners also widely recognize that ACs are useful for documenting assurance arguments in a structured and systematic manner, for communicating with interest holders and decision makers, and evaluating assurance-related evidence. Some practitioners view ACs as ``live'' and holistic models of assurance that evolve alongside the system being developed.

\paragraph{Expression of ACs.} Structured notations (e.g., GSN, CAE, EA, SACM) are now widely used by practitioners to express assurance arguments and are generally favoured as a means of organizing assurance-related information. However, some concerns exist related to the complexity of structured notations distracting from the essential purpose of ACs as a means to capture clear assurance arguments. Further, despite wide use of structured notations, narrative expression of ACs is still common, with many practitioners having experience with both structured notations and narratives.

\paragraph{Methods for Assessing Confidence in ACs.} Review is a very common\footnote{We hesitate to say that review is ``universally'' used because of our modest sample in this study, but it is worth noting that all participants identified review as a method.} means of assessing and increasing confidence in ACs, with some practitioners also using systematic review methods to ``interrogate'' an argument. Checklists derived from standards, guidance documents, organizational practices, and experience are used as review criteria to confirm all necessary topics have been addressed. Despite the clear importance of review, there might be an over-dependence on this as the only confidence assessment method, suggesting practitioners might value complementary methods. Dialectic argumentation (i.e., ``defeaters'' or ``challenges'') is also widely used as a means of mitigating bias in arguments. However, there are open methodological questions about the use of dialectic methods such as the absence of clear ``stopping'' criteria, handling of resolved defeaters, and distinguishing between genuine argument defeaters versus claims. While many practitioners are aware of quantitative methods for confidence assessment (e.g., BBNs, DST), they have seen very limited use for real-world ACs. Further, many practitioners have negative opinions of quantitative methods due to: a lack of confidence that the methods produce trustworthy results; loss of information when confidence is expressed as a number; difficulty explaining or justifying assessment results to interest holders or regulators; and volume of inputs required to use the method.

\paragraph{Uses of Confidence Assessment.} The results from confidence assessments are generally used to communicate to interest holders, understand the qualitative risk associated with the system, and improve the quality of an ACs argument or supporting evidence.

\paragraph{Barriers to Performing Confidence Assessment.} There are many barriers or challenges inhibiting the use of CAMs. First, these activities are perceived as adding to the workload of assurance experts, especially for qualitative methods that require a human to provide many inputs as part of the assessment. Second, there is also a lack of accessible guidance and training material on the theoretical basis for CAMs and how to select and apply in practice. Third, confidence assessment is seen as a subjective activity that produces variable results that can be difficult to interpret. Fourth, quantitative methods are relatively immature and there is a lack of empirical evidence to show they are trustworthy. Finally, the overall adoption of ACs as a system assurance method remains a limiting factor to practitioners applying AC confidence assessment: without an AC confidence cannot be assessed.

\subsection{Considerations for Future Research}

Given the importance of ACs in assuring modern systems, practitioners also recognize that it is important to perform confidence increasing or assessment activities and that they often have a positive impact on the project. However, there are a number of barriers that make it difficult for practitioners to use these methods in their industrial work. Further, there is a general sense from participants in our study that academic research on ACs has departed from the needs of practitioners. Therefore, there is a need for researchers to consider practitioners' needs when developing or improving CAMs. With this in mind, we offer five considerations for researchers working on either qualitative or quantitative AC confidence assessment:

\begin{enumerate}

    \item \bb{Connect.} Practitioners have established methods of assessing or increasing confidence in ACs, such as review, using checklists, or dialectic arguments. While not without their own challenges, these methods are already trusted by practitioners and are part of their AC practice. Researchers developing new (or tailoring existing) methods for confidence assessment should consider how their methods connect with these established practices. For instance, a researcher developing a new quantitative method should consider a use case where the method is used during initial creation of the AC and also during review of the AC.

    \item \bb{Crystallize.} One of the barriers to practitioners applying CAMs is a lack of a standardized and systematic method and procedure. Some methods have a well-developed theoretical basis, but then fail to address the practical aspects of application. Researchers should focus on crystallizing and maturing methods by formulating them as part of ``engineering procedures'' (or similar) that address topics such as: method applicability criteria, analysis completeness criteria, metrics for assessment progress, and criteria for interpreting results.
    
    \item \bb{Communicate.} Given the role of ACs as tools for communication with (potentially non-expert) interest holders, it is important to recognize that the results of confidence assessment will also be used as part of these communications. Researchers should consider how the results of methods will be viewed by interest holders and used to make decisions about critical systems, and what can be done from a methodological perspective to ensure that results are unambiguously interpreted.

    \item \bb{Confirm.} It is apparent that practitioners do not trust novel CAMs, especially quantitative ones. An important step towards gaining this trust is to confirm (i.e., validate) that methods have desirable characteristics such as repeatability, intuitiveness\footnote{By ``intuitiveness'' we mean whether the method produces results that align with experts' intuition about confidence in an AC or argument.}, and understandability. Validation should consist of both at-scale case studies and controlled experiments. Validation studies should also report on the level of effort required to apply the method to help practitioners estimate their own effort. It is worth noting that Graydon and Holloway also made a similar recommendation \cite{graydon2017}.
    
    \item \bb{Curate.} A key barrier identified by practitioners was a lack of accessible and actionable guidance or training materials on CAMs. In some cases, such guidance exists, but is not widely known to practitioners, while in other cases the only guidance available is the seminal publication on the method. As part of translating results on new or existing methods from research into practice, researchers should curate guidance and training materials that describe foundational theory in an approachable way, present methods in practical terms, and are widely available to the practice community.

\end{enumerate}

\subsection{Limitations and Threats to Validity}

This sub-section describes limitations and threats to the validity for the results from the present study.

First, this study used a grounded theory method aimed at understanding current practice. Through interviews and open coding analysis we were able to build up a detailed understanding of how practitioners think about confidence in ACs. The result is a ``theory'' (grounded in data) about current practice. While diversity in the participants helps improve the likelihood that this theory will generalize to describe the current practice for a wider sample of AC practitioners, we have not empirically confirmed this fact. That is, ours was a ``theory building'' rather than a ``theory validation'' exercise. To fully validate our results, future work must generate hypotheses based on the theory and then validate them using another study, which might consist of more interviews or perhaps a questionnaire.

Second, participants were recruited for this study using ``convenience sampling'' from the authors' professional networks and the sample is relatively small ($n=19$). It is possible that our sample is not representative of the larger AC practice community. To partially mitigate bias that our sample might have introduced, we sought practitioners with a wide range of industry experiences, and also aimed to include second (and even third) degree connections from the authors' professional networks. Nonetheless, our sampling method precludes us from making statistical inferences about whether our results are representative of the whole population of AC practitioners. Further, it is worth noting that statistical sampling methods, which are required for generalization, are impractical to apply for this population since there is no way to define a sampling frame from which to draw samples \cite{amir2018}. For instance, we could not have enumerated the complete population of AC practitioners from which to select a sample of participants for our study. 

Third, while the structured interview method provide an opportunity to collect very detailed data, it might also limit the completeness of the results. During interviews, it is possible that practitioners described their use of AC confidence assessment methods based on whatever was ``on their mind'' without considering a comprehensive list of potential answers. While the interview questions could have been more specific (e.g., ``have you used method X''), this could have also impacted the results by leading participants towards a pre-determined answer. Since our objective was one of theory building, not theory validation, we aired on the side of open-ended questions. It is possible that using more multiple choice questions might have produced different results.

Fourth, the target population for this study was AC practitioners, i.e., individuals that prepare ACs for real-world systems. As a result, the inclusion criteria for the study required that participants have experience preparing an ``assurance case''. For instance, at least two of the candidates who responded to the study's invitation did not meet this criterion, though they did have experience working on assurance for high-risk systems. It is likely that there is a population of practitioners who do not work with ACs and as a result use different methods for gaining confidence in their work products, which were not captured by this study.

\section{Conclusion}\label{sec:conclusion}

CAMs are of interest for practitioners focused on assessing or increasing confidence in ACs they have prepared or are managing. This paper surveyed CAMs from the literature, reported on a series of structured interviews with practitioners aimed at understanding the current industrial practice around ACs and confidence assessment, and identified considerations for researchers developing CAMs.

While many CAMs exist, there appears to be a gap between the needs and interests of practitioners and the methods described in the literature. Practitioners in this study were critical of quantitative CAMs, and favoured qualitative CAMs that emphasize (peer or independent) review, dialectic argumentation, as well as using checklists as criteria to guide reviews. Practitioners also identified barriers to performing CAMs, including additional effort, lack of accessible guidance on CAMs, subjectivity of methods, and a lack of trust in quantitative methods.

Going forward, we recommend that researchers working on CAMs consider five aspects. CAMs should \bb{connect} with existing practices trusted by practitioners. The importance of CAMs as a vehicle to \bb{communicate} with interest holders and decision makers should be recognized. The details of CAMs need to be \bb{crystallized} by include clear direction for their use. Concrete and accessible guidance should be \bb{curated} and made available to practitioners. Finally, further work is required to \bb{confirm} that CAMs are trustworthy before they can be used to influence decisions about critical systems.






\bibliographystyle{plain} 
\bibliography{refs} 

\begin{thebibliography}{10}

\bibitem{en50126}
{Railway} {Applications} - {The} {Specification} and {Demonstration} of
  {Reliability}, {Availability}, {Maintainability} and {Safety} ({RAMS}) -
  {Part} 1: {Generic} {RAMS} {Process}.
\newblock Standard {EN} 50126, European Committee for Electrotechnical
  Standardization, 2017.

\bibitem{iso26262}
{Road} {vehicles} - {Functional} safety.
\newblock Standard ISO 26262, International Organization for Standardization,
  2018.

\bibitem{iso21434}
{Road} vehicles - {Cybersecurity} engineering.
\newblock Standard ISO 21434, International Organization for Standardization,
  2021.

\bibitem{sacm}
Structured {Assurance} {Case} {Metamodel} ({SCAM}).
\newblock Standard, Object Management Group, 2021.

\bibitem{ul4600}
{Standard} for {Evaluation} of {Autonomous} {Products}.
\newblock Standard UL 4600, Underwritter Laboratories, 2022.

\bibitem{almendra2022}
Camilo Almendra, Carla Silva, Luiz Eduardo~G. Martins, and Johnny Marques.
\newblock How assurance case development and requirements engineering
  interplay: a study with practitioners.
\newblock {\em Requirements Engineering}, 27:273--292, 2022.

\bibitem{amir2018}
Bilal Amir and Paul Ralph.
\newblock There is no random sampling in software engineering research.
\newblock In {\em Proceedings of the 40th {International} {Conference} on
  {Software} {Engineering}: {Companion} {Proceeedings}}, pages 344--345.
  Association for Computing Machinery, 2018.

\bibitem{asaadi2020}
Erfan Asaadi, Ewen Denney, Jonathan Menzies, Ganesh~J. Pai, and Dimo Petroff.
\newblock Dynamic {Assurance} {Cases}: {A} {Pathway} to {Trusted} {Autonomy}.
\newblock {\em Computer}, 53(12):35--46, 2020.

\bibitem{gsn}
{Assurance Case Working Group}.
\newblock Goal {Structuring} {Notation} {Community Standard} {(Version 3)}.
\newblock Standard, Safety Critical Systems Club, 2021.

\bibitem{ayoub2013}
Anaheed Ayoub, Jian Chang, Oleg Sokolsky, and Insup Lee.
\newblock Assessing the {Overall} {Sufficiency} of {Safety} {Arguments}.
\newblock In {\em 21st Safety-Critical Systems Symposium (SSS'13)}, pages
  127--144. Safety-Critical Systems Club, 2013.

\bibitem{bloomfield2023}
Robin Bloomfield and John Rushby.
\newblock {Assessing Confidence} with {Assurance 2.0}.
\newblock Technical Report SRI-CSL-2022-02, SRI International, 2022.

\bibitem{calinescu2018}
Radu Calinescu, Danny Weyns, Simos Gerasimou, Muhammad~Usman Iftikhar, Ibrahim
  Habli, and Tim Kelly.
\newblock Engineering trustworthy self-adaptive software with dynamic assurance
  cases.
\newblock {\em {IEEE} Transactions on Software Engineering}, 44(11):1039--1069,
  2018.

\bibitem{cheng2018}
Jinghui Cheng, Micayla Goodrum, Ronald Metoyer, and Jane Cleland-Huang.
\newblock How do practitioners perceive assurance cases in safety-critical
  software systems?
\newblock In {\em 11th {International} {Workshop} on {Cooperative} and {Human}
  {Aspects} of {Software} {Engineering}}, pages 57--60. Association for
  Computing Machinery, 2018.

\bibitem{cyra2011}
Lukasz Cyra and Janusz Górski.
\newblock Support for argument structures review and assessment.
\newblock {\em Reliability Engineering \& System Safety}, 96(1):26--37, 2011.

\bibitem{denney2015}
E.~Denney, G.~Pai, and I.~Habli.
\newblock Dynamic {Safety} {Cases} for {Through}-{Life} {Safety} {Assurance}.
\newblock In {\em 2015 {IEEE}/{ACM} 37th {IEEE} {International} {Conference} on
  {Software} {Engineering}}, volume~2, pages 587--590, 2015.

\bibitem{denney2011}
Ewen Denney, Ganesh Pai, and Ibrahim Habli.
\newblock Towards {Measurement} of {Confidence} in {Safety} {Cases}.
\newblock In {\em 2011 {International} {Symposium} on {Empirical} {Software}
  {Engineering} and {Measurement}}, pages 380--383, 2011.

\bibitem{diemert2020}
S.~Diemert and J.~Joyce.
\newblock Eliminative {Argumentation} for {Arguing} {System} {Safety} - {A}
  {Practitioner}'s {Experience}.
\newblock In {\em 2020 {IEEE} {International} {Systems} {Conference}
  ({SysCon})}, pages 1--7, 2020.

\bibitem{diemert2024}
Simon Diemert, Laure Millet, Jeff Joyce, and Jens~H. Weber.
\newblock Including {Defeaters} in {Quantitative} {Confidence} {Assessments}
  for {Assurance} {Cases}.
\newblock In Andrea Ceccarelli, Mario Trapp, Andrea Bondavalli, Erwin
  Schoitsch, Barbara Gallina, and Friedemann Bitsch, editors, {\em Computer
  {Safety}, {Reliability}, and {Security}. {SafeCOMP} 2024 {Workshops}}, volume
  14989 of {\em Lecture Notes in Computer Science}, pages 239--250. Springer,
  2024.

\bibitem{doss2016}
O.~Doss and T.~P. Kelly.
\newblock Challenges and {Opportunities} in {Agile} {Development} in {Safety}
  {Critical} {Systems}: {A} {Survey}.
\newblock {\em ACM SIGSOFT Software Engineering Notes}, 41(2):30--31, 2016.

\bibitem{duan2015}
Lian Duan, Sanjai Rayadurgam, Mats P.~E. Heimdahl, Oleg Sokolsky, and Insup
  Lee.
\newblock Representing {Confidence} in {Assurance} {Case} {Evidence}.
\newblock In Floor Koornneef and Coen van Gulijk, editors, {\em Computer
  {Safety}, {Reliability}, and {Security} ({SAFECOMP} 2015)}, volume 9338 of
  {\em Lecture {Notes} in {Computer} {Science}}, pages 15--26. Springer, 2015.

\bibitem{dubois1988}
Didier Dubois and Henri Prade.
\newblock {\em Possibility {Theory}: {An} {Approach} to {Computerized}
  {Processing} of {Uncertainty}}.
\newblock Plenum Press, New York, NY, USA, 1988.

\bibitem{easterbrook2008}
Steve Easterbrook, Janice Singer, Margaret-Anne Storey, and Daniela Damian.
\newblock Selecting {Empirical} {Methods} for {Software} {Engineering}
  {Research}.
\newblock In Forrest Shull, Janice Singer, and Dag I.~K. Sjøberg, editors,
  {\em Guide to Advanced Empirical Software Engineering}, pages 285--311.
  Springer, 2008.

\bibitem{fenn2024}
Jane Fenn, Richard Hawkins, and Mark Nicholson.
\newblock A {New} {Approach} to {Creating} {Clear} {Operational} {Safety}
  {Arguments}.
\newblock In Andrea Ceccarelli, Mario Trapp, Andrea Bondavalli, Erwin
  Schoitsch, Barbara Gallina, and Friedemann Bitsch, editors, {\em Computer
  {Safety}, {Reliability}, and {Security}. {SafeCOMP} 2024 {Workshops}}, pages
  227--238, Cham, 2024. Springer.

\bibitem{onr2019}
Office for Nuclear~Regulation.
\newblock {The} {Purpose}, {Scope}, and {Content} of {Safety} {Cases}.
\newblock Technical Report NS-TAST-GD-051, Government of the United Kingdom,
  2019.

\bibitem{aami2019}
Association for the Advancement~of Medical~Instrumentation.
\newblock {AAMI} {TIR38:2019} - {Medical} {Device} {Safety} {Assurance} {Case}
  {Guidance}, 2019.

\bibitem{glaser1999}
Barney Glaser and Anselm Strauss.
\newblock {\em Discovery of Grounded Theory: Strategies for Qualitative
  Research}.
\newblock Routledge, 1st edition, 1999.

\bibitem{goodenough2013}
John~B. Goodenough, Charles~B. Weinstock, and Ari~Z. Klein.
\newblock Eliminative {Induction}: {A} basis for arguing system confidence.
\newblock In {\em 35th {International} {Conference} on {Software}
  {Engineering}}, pages 1161--1164, 2013.

\bibitem{goodenough2015}
John~B. Goodenough, Charles~B. Weinstock, and Ari~Z. Klein.
\newblock Eliminative {Argumentation}: A {Basis} for {Arguing} {Confidence} in
  {System} {Properties}.
\newblock Technical report, Carnegie Mellon University - Software Engineering
  Institute, 2015.

\bibitem{graydon2016}
Patrick~J. Graydon.
\newblock Defining baconian probability for use in assurance argumentation.
\newblock Technical Report {NASA}/{TM}-2016-219341, National Aeronautics and
  Space Adminstration ({NASA}), 2016.

\bibitem{graydon2017}
Patrick~J. Graydon and C.~Michael Holloway.
\newblock An investigation of proposed techniques for quantifying confidence in
  assurance arguments.
\newblock {\em Safety Science}, 92:53--65, 2017.

\bibitem{acwg2021}
Assurance Case~Working Group.
\newblock {Assurance} {Case} {Guidance} - {Challenges}, {Common} {Issues} and
  {Good} {Practice} (version 1.1).
\newblock Guidance, Safety Critical Systems Club, 2021.

\bibitem{nimrod}
Charles Haddon-Cav.
\newblock The {Nimrod} {Review}.
\newblock Independent {Review}, House of Commons of the United Kingdom, 2009.

\bibitem{hawkins2011}
Richard Hawkins, Tim Kelly, John Knight, and Patrick Graydon.
\newblock A {New Approach} to {Creating} {Clear} {Safety Arguments}.
\newblock In {\em Advances in Systems Safety}, pages 3--23. Springer, 2011.

\bibitem{herd2024}
Benjamin Herd, João-Vitor Zacchi, and Simon Burton.
\newblock A {Deductive} {Approach} to {Safety} {Assurance}: {Formalising}
  {Safety} {Contracts} with {Subjective} {Logic}.
\newblock In Andrea Ceccarelli, Mario Trapp, Andrea Bondavalli, Erwin
  Schoitsch, Barbara Gallina, and Friedemann Bitsch, editors, {\em Computer
  {Safety}, {Reliability}, and {Security}. {SAFECOMP} 2024 {Workshops}}, volume
  14989 of {\em Lecture {Notes} in {Computer} {Science}}, pages 213--226.
  Springer, 2024.

\bibitem{hobbs2019}
Chris Hobbs.
\newblock Experience with {Assurance} {Case} {Preparation}.
\newblock Experience report, BlackBerry QNX, 2019.

\bibitem{hobbs2012}
Chris Hobbs and Martin Lloyd.
\newblock The {Application} of {Bayesian Belief Networks} to {Assurance Case}
  {Preparation}.
\newblock In C.~Dale and T.~Anderson, editors, {\em Achieving Systems Safety},
  pages 159--176. Springer, 2012.

\bibitem{holloway2019}
C.~Michael Holloway.
\newblock Understanding the {Overarching} {Properties}.
\newblock Technical Report NASA/TM-2019-220292, National Aeronautics and Space
  Administration (NASA), 2019.

\bibitem{fan}
C.~Michael Holloway.
\newblock The {Friendly} {Argument} {Notation} ({FAN}): 2023 {Version}.
\newblock Technical Report NASA/TM-20230004423, National Aeronautics and Space
  Administration (NASA), 2023.

\bibitem{holloway2021}
C.~Michael Holloway and Kimberly~S. Wasson.
\newblock A {Primer} on {Argument} {Assessment}.
\newblock Technical report, National Aeronautics and Space Administration
  (NASA) and Joby Aviation, 2021.

\bibitem{idmessaoud2024}
Yassir Idmessaoud, Didier Dubois, and Jérémie Guiochet.
\newblock Confidence assessment in safety argument structure - {Quantitative}
  vs. qualitative approaches.
\newblock {\em International Journal of Approximate Reasoning}, 165:109100,
  2024.

\bibitem{offshore2005}
UK~Statutory Instruments.
\newblock {The} {Offshore} {Installations} ({Safety} {Case}) {Regulations}
  2005.
\newblock Regulation, Government of the United Kingdom, 2005.

\bibitem{josang2016}
Audun Jøsang.
\newblock {\em Subjective {Logic}}.
\newblock Artificial {Intelligence}: {Foundations}, {Theory}, and {Algorithms}.
  Springer, 2016.

\bibitem{kasunic2005}
Mark Kasunic.
\newblock Designing an {Effective} {Survey}.
\newblock Technical Report CMU/SEI-2005-HB-004, Carnegie Mellon University -
  Software Engineering Institute, 2005.

\bibitem{kelly}
Tim~P. Kelly.
\newblock {\em {Arguing} {Safety} - {A} {Systematic} {Approach} to {Safety}
  {Case} {Management}}.
\newblock {PhD} thesis, 1998.

\bibitem{kohli2020}
Puneet Kohli and Anjali Chadha.
\newblock Enabling {Pedestrian} {Safety} {Using} {Computer} {Vision}
  {Techniques}: {A} {Case} {Study} of the 2018 {Uber} {Inc}. {Self}-driving
  {Car} {Crash}.
\newblock In Kohei Arai and Rahul Bhatia, editors, {\em Advances in
  {Information} and {Communication}}, Lecture {Notes} in {Networks} and
  {Systems}, pages 261--279. Springer International Publishing, 2020.

\bibitem{koopman2024}
Philip Koopman.
\newblock Anatomy of a {Robotaxi} {Crash}: {Lessons} from the {Cruise}
  {Pedestrian} {Dragging} {Mishap}.
\newblock In Andrea Ceccarelli, Mario Trapp, Andrea Bondavalli, and Friedemann
  Bitsch, editors, {\em Computer {Safety}, {Reliability}, and {Security}
  ({SAFECOMP}) 2024}, volume 14988 of {\em Lecture {Notes} in {Computer} and
  {Science}}, pages 119--133. Springer, 2024.

\bibitem{leveson2011}
Nancy~G. Leveson.
\newblock The {Use} of {Safety} {Cases} in {Certification} and {Regulation}.
\newblock \url{https://dspace.mit.edu/handle/1721.1/102833}, 2011.

\bibitem{linaker2015}
Johan Linaker, Sardar~Muhammad Sulaman, Rafael Maiani~de Mello, and Martin
  Host.
\newblock Guidelines for {Conducting} {Surveys} in {Software} {Engineering}.
\newblock Guideline, Lunds University, 2015.

\bibitem{millet2023}
Laure Millet, Simon Diemert, Chris Rees, Torin Viger, Marsha Chechik, Claudio
  Menghi, and Jeffrey Joyce.
\newblock {Assurance Case} {Arguments} in the {Large}: The {CERN} {LHC}
  {Machine Protection System}.
\newblock In {\em {Computer Safety}, {Reliability}, and {Security}. {SAFECOMP}
  2023}, volume 14181 of {\em Lecture Notes in Computer Science}, pages 3--10.
  Springer, 2023.

\bibitem{nair2015}
Sunil Nair, Jose~Luis de~la Vara, Mehrdad Sabetzadeh, and Davide Falessi.
\newblock Evidence management for compliance of critical systems with safety
  standards: {A} survey on the state of practice.
\newblock {\em Information and Software Technology}, 60:1--15, 2015.

\bibitem{defstan56}
United Kingdom~Ministry of~Defence.
\newblock Defence {Standard} 00-056 - {Safety} {Management} {Requirements} for
  {Defence} {Systems}.
\newblock Standard Def Stan 00-056, 2023.

\bibitem{oh2022}
Chanwook Oh, Nikhil Naik, Zamira Daw, Timothy~E. Wang, and Pierluigi Nuzzo.
\newblock {ARACHNE}: {Automated} {Validation} of {Assurance} {Cases} with
  {Stochastic} {Contract} {Networks}.
\newblock In {\em {International} {Conference} on {Computer} {Safety},
  {Reliability} and {Security}, {SafeCOMP} 2022}, volume 13414 of {\em Lecture
  {Notes} in {Computer} {Science}}, pages 65--81. Springer, 2022.

\bibitem{rees2023}
Chris Rees, Mateo Delgado, Rolf Lippet, Jeffrey Joyce, Simon Diemert, Claudio
  Menghi, Torin Viger, Marsha Chechik, Jan Uythoven, Markus Zerlauth, and Lukas
  Felsberfer.
\newblock Assessing the {Usefulness} of {Assurance} {Cases}: an {Experience}
  with the {CERN} {Large} {Hadron} {Collider}.
\newblock Experience Report CERN-ACC-2023-0002, European Organization for
  Nuclear Research (CERN), 2023.

\bibitem{ryan2024}
Philippa Ryan, Sepeedeh Shahbeigi, Jie Zou, Ioannis Stefanakos, and John
  Molloy.
\newblock A {Dynamic} {Assurance} {Framework} for an {Autonomous} {Survey}
  {Drone}.
\newblock In Andrea Ceccarelli, Mario Trapp, Andrea Bondavalli, and Friedemann
  Bitsch, editors, {\em Computer {Safety}, {Reliability}, and {Security}.
  {SAFECOMP} 2024}, pages 285--299. Springer, 2024.

\bibitem{seaman1999}
C.B. Seaman.
\newblock Qualitative methods in empirical studies of software engineering.
\newblock {\em {IEEE} Transactions on Software Engineering}, 25(4):557--572,
  1999.

\bibitem{shafer1976}
Glenn Shafer.
\newblock {\em A {Mathematical} {Theory} of {Evidence}}.
\newblock Princeton University Press, 1976.

\bibitem{shahandashti2024}
Kimya~Khakzad Shahandashti, Alvine~B. Belle, Timothy~C. Lethbridge, Oluwafemi
  Odu, and Mithila Sivakumar.
\newblock A {PRISMA}-driven systematic mapping study on system assurance
  weakeners.
\newblock {\em Information and Software Technology}, 175:107526, 2024.

\bibitem{toulmin}
Stephen~E. Toulmin.
\newblock {\em The {Uses} of {Argument}}.
\newblock Cambridge University Press, 2nd edition, 2003.

\bibitem{dagstuhl2024}
Elena Troubitsyna, Ignacio~J. Alvarez, Philip Koopman, and Mario Trapp.
\newblock Methods and {Tools} for the {Engineering} and {Assurance} of {Safe}
  {Autonomous} {Systems} ({Dagstuhl} {Seminar} 24151).
\newblock In {\em Dagstuhl Reports}, volume~14, pages 23--41, 2024.

\bibitem{torner2008}
Fredrik Törner and Peter Öhman.
\newblock Automotive {Safety} {Case} {A} {Qualitative} {Case} {Study} of
  {Drivers}, {Usages}, and {Issues}.
\newblock In {\em 11th {IEEE} {High} {Assurance} {Systems} {Engineering}
  {Symposium} ({HASE})}, pages 313--322, 2008.

\bibitem{wang2019}
Rui Wang, Jérémie Guiochet, Gilles Motet, and Walter Schön.
\newblock Safety case confidence propagation based on {Dempster}-{Shafer}
  theory.
\newblock {\em International Journal of Approximate Reasoning}, 107:46--64,
  April 2019.

\bibitem{wasson2022}
Kimberly~S. Wasson and C.~Michael Holloway.
\newblock An {Introduction} to {Constructing} and {Assessing} {Overarching}
  {Properties} {Related} {Arguments} ({OPRAs}): {Version} 1.0.
\newblock Technical report, {Joby Aviation} and National Aeronautics and Space
  Administration (NASA), 2022.

\bibitem{yuan2017}
Chunchun Yuan, Ji~Wu, Chao Liu, and {and}~Haiyan Yang.
\newblock A {Subjective} {Logic}-{Based} {Approach} for {Assessing}
  {Confidence} in {Assurance} {Case}.
\newblock {\em International Journal of Performability Engineering}, 13(6):807,
  2017.

\bibitem{zadeh1965}
Lotfi Zadeh.
\newblock Fuzzy {Sets}.
\newblock {\em Information and Control}, 8(3):338--353, 1965.

\bibitem{zhao2012}
Xingyu Zhao, Dajian Zhang, Minyan Lu, and Fuping Zeng.
\newblock A {New} {approach} to {Assessment} of {Confidence} in {Assurance}
  {Cases}.
\newblock In {\em International Conference on Computer Safety, Reliability and
  Security (SafeCOMP) 2012}, volume 7613 of {\em Lecture Notes in Computer
  Science}, pages 79--91. Springer, 2012.

\end{thebibliography}

\appendix
\newpage
\section{Interview Questions}\label{appendix:questions}
The interview consisted of the following questions:

\begin{enumerate}

    \item {\color{red}\bb{[SCREENING]}} How many years of professional experience do you have? Both in general and working in systems or software assurance? \it{Participant must report some amount of professional experience}.

    \item What industry or industries do you have experience working in? What industries have you prepared ACs in?

    \item {\color{red}\bb{[SCREENING]}} Please describe your role as it relates to the preparation or development of ACs.

    \item What are the quality attributes being assured in the ACs you have worked on? For example, a ``safety case''.

    \item How are the ACs that you have worked on expressed or captured?

    \item How have you evaluated or assessed your degree of belief or confidence in an AC? What methods, techniques, or activities have you used?

    \item Have you ever used a quantitative method for assessing confidence as part of an real-world project? If so, what was your opinion of the method and the outcome?
    
    \item Given that you have performed some type of confidence assessment, how have you used the results?

    \item {\color{blue}\bb{[MUTIPLE CHOICE]}} What impact did the methods, techniques, or activities you used to assess confidence have on the ACs you have worked on?
    
        \begin{itemize}
            \item Positive Impact
            \item Neutral Impact
            \item Negative Impact
        \end{itemize}

    \item What challenges or barriers do you think exist, or have you experienced, to conducting confidence assessments for ACs in practice?

    \item {\color{blue}\bb{[MUTIPLE CHOICE]}} Regardless of what you or your organization have done in the past, do you believe a dedicated confidence assessment activity is important? Should it be:

        \begin{itemize}
            \item Mandatory
            \item Recommended
            \item Optional
            \item Discouraged
        \end{itemize}

\end{enumerate}

\section{Detailed Code Book}\label{appendix:codes}

\begin{landscape}

\newcommand{\lightrule}{\arrayrulecolor{black!30}\cmidrule{2-4}\arrayrulecolor{black}}
\newcommand{\cellCatWidth}{3cm}
{
\footnotesize
\def\arraystretch{1.25}
\begin{longtable}{p{\cellCatWidth}p{2.5cm}p{12cm}c}

        \toprule

        \bb{Category} & \bb{Code} & \bb{Code Definition} & \bb{Participants} \\ \midrule
        \endhead

        \\
        \multicolumn{4}{c}{\it{Continued on next page...}}\\
        \endfoot

        \bottomrule
        \endlastfoot

        \multirow{10}{\cellCatWidth}[-5cm]{\bi{Motivation} for preparing ACs} 

            & Capture Arguments 
                & A motivation for preparing an AC is to document assurance arguments in a structured or systematic manner. 
                & 12 \\ \lightrule 

            & Compliance
                & A motivation for preparing an AC is to satisfy, or check compliance with, technical standards, regulatory requirements, or contractual obligations.
                & 12 \\ \lightrule

            & Live Model of Assurance
                & A motivation for preparing an AC is to have a ``live'' model of an AC that is updated periodically, perhaps evolving alongside the system(s) it assures.
                & 6 \\ \lightrule

            & Evidence Adequacy 
                & A motivation for preparing an AC is to determine if the available evidence is adequate to support the assurance argument(s), including the possibility of making a ``self-checklist'' of the required evidence.
                & 4 \\ \lightrule

            & Holistic Assurance
                & A motivation for preparing an AC is to develop an overall view of assurance, across multiple parts of an organization or a system, especially where the different parts might have different objectives (i.e., ACs help address assurance during system integration).
                & 4 \\ \lightrule

            & Lack of Guidance 
                & A motivation for preparing an AC is that there is a lack of prescriptive guidance on how to assure novel or complex systems, so it is necessary to make an argument about why you (or your organization) believes a system meets its stated objective(s).
                & 3 \\ \lightrule

            & Risk Management 
                & A motivation for preparing an AC is to understand and manage risk associated with a system, including hazard identification, risk analysis, and mitigations.
                & 3 \\ \lightrule

            & Gap Analysis 
                & A motivation for preparing an AC is to discover or understand ``gaps'' in (previously implicit) assurance arguments or assurance-related activities.
                & $\leq 2$ \\ \lightrule

            & Complexity of Systems
                & A motivation for preparing an AC is to manage the increasing complexity of systems, which now require more sophisticated assurance arguments. 
                & $\leq 2$ \\ \lightrule

            & Deployment Decision 
                & A motivation for preparing an AC is to help decide to deploy a system, based on the argument(s) and evidence presented in the AC. 
                & $\leq 2$ \\ \midrule

        \multirow{7}{\cellCatWidth}{\bi{Expression} of ACs} 

            & GSN 
                & Participant has used Goal Structuring Notation (GSN), or a GSN-like notation, to express assurance arguments.
                & 15 \\ \lightrule 

            & Narratives 
                & Participant has used narrative, report, or ``essay'' formats to express assurance arguments.
                & 13 \\ \lightrule 

            & CAE 
                & Participant has used Claims-Argument-Evidence (CAE) notation ot express assurance arguments.
                & 6 \\ \lightrule 

            & EA 
                & Participant has used Eliminative Argumentation (EA) notation to express assurance arguments.
                & 5 \\ \lightrule 

            & Tabular 
                & Participant has used a tabular format (e.g., a spreadsheet) to express assurance arguments.
                & $\leq 2$ \\ \lightrule 

            & FAN 
                & Participant has used the Friendly Argument Notation (FAN) to express assurance arguments.
                & $\leq 2$ \\ \lightrule 

            & SACM 
                & Participant has used the Structured Assurance Case Meta-Model (SACM)  to express assurance arguments.
                & $\leq 2$ \\ \midrule 

        \multirow{7}{\cellCatWidth}{Quality \bi{Attributes} for ACs} 

            & Safety 
                & Participant has prepared an AC where safety is the main quality attribute being considered. 
                & 16 \\ \lightrule 

            & Security 
                & Participant has prepared an AC where security is the main quality attribute being considered. 
                & 4 \\ \lightrule 

            & Mixing Safety \& Security
                & Participant has prepared an AC where a combination of safety and security is being considered. 
                & $\leq 2$ \\ \lightrule 

            & Mixing Many Attributes 
                & Participant has prepared an AC where a combination of many quality attributes (safety, security, reliability, availability, fitness-for-purpose, etc.) are considered.
                & 11 \\ \midrule 

        \multirow{7}{\cellCatWidth}{\bi{Methods} for Confidence Assessment} 

            & Review 
                & Participant has using peer or independent review as a means of assessing or increasing confidence in an AC, including ``recurring reviews'' at key milestones
                & 19 \\ \lightrule 

            & Defeaters 
                & Participant has experience using dialectic approaches (i.e., defeaters, doubts, challenges, etc.) to increase confidence in an AC.
                & 12 \\ \lightrule 

            & Checklists 
                & Participant has experience assessing confidence in an AC by checking it against external authorities such as technical standards or guidelines (``did we think of that?'').
                & 9 \\ \lightrule 

            & Personnel
                & Participant has experience assessing confidence in an AC by reasoning about the qualities of the personnel who worked on the AC.
                & 5 \\ \lightrule 

            & Interrogation
                & Participant has experience interrogating the argument structure (premises $\rightarrow$ conclusion), either using a systematic method (e.g., SPRY/iTest), guidewords, or informally as a means of self-organizing a review.
                & 5 \\ \lightrule 

            & Confidence Arguments
                & Participant has experience preparing supplementary/supporting confidence arguments focused on the confidence in the main argument (e.g., Assurance Claim Points).
                & 3 \\ \lightrule 

            & Counting Ratios
                & Participant has experience assessing confidence based on ratios derived from counting elements in the argument structures (e.g., number of defeaters v. claims).
                & 3 \\ \lightrule 

            & Assurance 2.0
                & Participant has experience applying the Assurance 2.0 method to assess and increase confidence in an AC.
                & $\leq 2$ \\ \lightrule 

            & BNNs
                & Participant has experience applying Bayesian Belief Networks to assess confidence in an AC.
                & $\leq 2$ \\ \lightrule 

            & Coarse Belief 
                & Participant has experience using coarse belief measures (e.g., ``gut feel'', ``on a scale from 0 - 10...'').
                & $\leq 2$\\ \lightrule 

            & Performance Indicators
                & Participant has experience using safety or key performance indicators, based on engineering or operational data, to assess confidence in an argument.
                & $\leq 2$ \\ \lightrule 

            & DST 
                & Participant has experience using Dempster-Shafer Theory based approaches for assessing confidence in an AC.
                & $\leq 2$ \\ \midrule 

        \multirow{7}{\cellCatWidth}{\bi{Barriers} to Conducting Confidence Assessment} 

            & More Work 
                & Participant identified that conducting a confidence assessment requires additional effort/work, which might impact project schedule or budgets.
                & 13 \\ \lightrule 

            & Inadequate Guidance
                & Participant identified that guidance or training on confidence assessments is missing, inadequate, or otherwise not accessible to practitioners, so practitioners don't use the methods.
                & 12 \\ \lightrule 

            & Interpreting Results
                & Participant identified that it can be difficult to interpret the results produced by a confidence assessment method and make decisions based on them.
                & 10 \\ \lightrule 

            & Confidence in Confidence
                & Participant identified a meta-problem exists where an analyst might ask if they trust the assessed level of confidence (e.g., stopping rule for defeaters).
                & 9 \\ \lightrule 

            & Subjective 
                &  Participant identified that confidence assessment method(s) are too subjective, relying on expert opinions, which might vary significantly.
                & 9 \\ \lightrule 

            & Immature 
                & Participant identified that confidence assessment method(s) are too immature to be used in practice or depended upon to make decisions about critical systems.
                & 8 \\ \lightrule 

            & Safety Culture
                & Participant identified that poor safety culture is a barrier to performing confidence assessments.
                & 7 \\ \lightrule 

            & Lack of Adoption
                & Participant identified that confidence assessment methods are not applicable because assurance cases are not adopted in some industries or organizations.
                & 6 \\ \lightrule 

            & Terminology
                & Participant identified that different methods and approaches have different terminology and conceptual models, which makes them difficult to apply for practitioners.
                & 3 \\ \lightrule 

            & Change Impact
                & Participant identified that it can be difficult to (re-)assess confidence when the argument changes. 
                & $\leq 2$ \\ \lightrule 

            & Tooling
                & Participant identified that there is a lack of mature tooling to support confidence assessment
 
                & $\leq 2$ \\ \lightrule 

            & Not Required
                & Participant identified that confidence assessments are not required by industry standards or practices, so they don't do them.
  
                & $\leq 2$ \\ \lightrule 

            & Not Integrated
                & Participant identified that confidence assessment method(s) are not deeply integrated into structured notations (e.g., GSN) and so can be difficult to use.
                & $\leq 2$ \\ \lightrule 

            & Project Management
                & It can be difficult to weave confidence assessments into project management/planning (e.g., too much/too early and the results are not useful, and too late and the results are overwhelming).
                & $\leq 2$ \\ \midrule 

        \multirow{7}{\cellCatWidth}{\bi{Uses} of Confidence Assessment Results} 

            & Communication with Stakeholders 
                & The participant has experience using the results of confidence assessment to communicate with project or system stakedholders. 
                & 14 \\ \lightrule 

            & Addressing Issues 
                & The participant has experience using the results from confidence assessment is used to address ``issues'' (or defects, problems, etc.) in the system or operations or in the AC itself.
                & 8 \\ \lightrule 

            & Understanding Risk
                & The participant has experience using the results of confidence assessment to help understand residual risk (quantitatively or qualitatively) of a system.
                & 8 \\ \lightrule 

            & Evaluate Argument 
                & The participant has experience using the results of confidence assessment to evaluate the argument structure, including whether conclusions are justified.
                & 7 \\ \lightrule 

            & Deployment Decision 
                & The participant has experience using the results of confidence assessment to make a system deployment decision.
                & 6 \\ \lightrule 

            & Evidence Evaluation 
                & The participant has experience using the result of confidence assessment to evaluate whether evidence is suitable for the argument. 
                & 6 \\ \lightrule 

            & Avoid Confirmation Bias 
                & The participant has experience using the results of confidence assessment help mitigate confirmation bias.
                & 5 \\ 

\end{longtable}
}

\end{landscape}

\end{document}